\documentclass[10pt,showpacs,floatfix,letterpaper,amsmath,amsfonts,amssymb,aps,pra,reprint,superscriptaddress]{revtex4-1}
\usepackage{color}
\usepackage{tensor}
\usepackage{graphicx}
\usepackage{amsmath}
\usepackage{epstopdf}
\usepackage{enumerate}
\usepackage[latin1]{inputenc}
\usepackage{bm}
\usepackage{xfrac}
\usepackage[caption=false]{subfig}

\usepackage[pdfpagemode=UseNone,pdfstartview=FitH,colorlinks=true,linkcolor=blue,citecolor=blue,urlcolor=blue]{hyperref}
\usepackage[all]{hypcap}

\newcommand{\ket}[1]{\lvert #1\rangle}           
\newcommand{\bra}[1]{\langle #1\rvert}           
\newcommand{\mean}[1]{\langle #1 \rangle}        
\newcommand{\be} {\begin{equation}}
\newcommand{\ee} {\end{equation}}
\def\d{\mathrm{d}}                               

\begin{document}
\title{Qubit measurement error from coupling with a detuned neighbor in circuit QED}
\author{Mostafa Khezri}
\email[email: ]{mostafa.khezri@email.ucr.edu}
\affiliation{Department of Electrical and Computer Engineering, University of California, Riverside, CA 92521, USA.}
\affiliation{Department of Physics, University of California, Riverside, CA 92521, USA.}
\author{Justin Dressel}
\affiliation{Department of Electrical and Computer Engineering, University of California, Riverside, CA 92521, USA.}
\author{Alexander N. Korotkov}
\affiliation{Department of Electrical and Computer Engineering, University of California, Riverside, CA 92521, USA.}
\date{\today}

\pacs{03.67.Lx, 85.25.-j}

\begin{abstract}
In modern circuit QED architectures, superconducting transmon qubits
are measured via the state-dependent phase and amplitude shift of a
microwave field leaking from a coupled resonator. Determining this
shift requires integrating the field quadratures for a nonzero
duration, which can permit unwanted concurrent evolution. Here we
investigate such dynamical degradation of the measurement fidelity
caused by a detuned neighboring qubit. We find that in realistic
parameter regimes, where the qubit ensemble-dephasing rate is slower
than the qubit-qubit detuning, the joint qubit-qubit eigenstates are
better discriminated by measurement than the bare states.
Furthermore, we show that when the resonator leaks much more slowly
than the qubit-qubit detuning, the measurement tracks the joint
eigenstates nearly adiabatically. However, the measurement process
also causes rare quantum jumps between the eigenstates. The rate of
these jumps becomes significant if the resonator decay is comparable
to or faster than the qubit-qubit detuning, thus significantly
degrading the measurement fidelity in a manner reminiscent of energy
relaxation processes.
\end{abstract}
\maketitle

\section{Introduction}\label{sec:intro}

Recent years have witnessed the rapid evolution of superconducting circuit QED technology for quantum computation \cite{Blais2004,Wallraff2004,Sun2014,Barends2014, Chow2014,Weber2014,Riste2013, Lin2014,Dewes2012}
(reviewed in \cite{Clarke2008,You2011,Devoret2013}). The most recent developments have converged on charge-insensitive designs, based on transmons \cite{Koch2007}, which can be dispersively measured with coupled microwave resonators. Multi-qubit chips based on these designs have recently demonstrated high-fidelity entangling gates \cite{Barends2014,Chow2014}, which are now nearing the gate fidelity thresholds necessary for implementing practical quantum error correction protocols \cite{bookNielsen2000,Kitaev1997,Fowler2012a}. Indeed, several groups have recently demonstrated bit-flip error correction in such multi-qubit superconducting processors \cite{Reed2012,Kelly2015,Corcoles2015,Riste2015}. With the gate fidelity reaching such unprecedented levels, it is now interesting to identify and address more subtle sources of error that can arise in such a multi-qubit environment, such as the effect of non-tunable qubit-qubit or qubit-bus coupling \cite{Galiautdinov2012} on the dispersive measurement fidelity.

Unlike the textbook projective measurements usually assumed in the quantum computing literature \cite{bookNielsen2000}, which involve instantaneous state collapse, realistic measurements occur over a nonzero duration of time. In the transmon-based circuits we consider here, each qubit is dispersively coupled to a pumped microwave resonator such that the leaked field is phase-shifted (and, in general, amplitude-shifted) by a qubit-state-dependent amount \cite{Blais2004}. The leaked field is then passed through an amplifier and mixed with a local oscillator to produce a noisy homodyne signal. This signal needs to be integrated until
the signal-to-noise ratio exceeds an acceptable discrimination threshold (more advanced signal processing techniques can moderately increase the measurement fidelity \cite{Gambetta2007}). For an isolated qubit, the increase of the signal-to-noise ratio by longer integration is limited by the energy relaxation (and excitation) processes. However, in circuits intended for quantum computation, the qubits will also be coupled to frequency-detuned neighbors, which may permit unwanted dynamics to additionally degrade the measurement fidelity. We wish to better understand the detailed dynamics of a realistic transmon qubit measurement, and minimize the dynamical measurement error that will arise from the coupling to a neighboring qubit (or bus, which plays a similar role).

In this paper, we demonstrate that in typical experimental parameter regimes, where the qubit ensemble-dephasing rate due to measurement is slower than the qubit-qubit detuning, dynamical measurement error always exists when distinguishing the bare energy states of coupled qubits. However, this measurement error can be decreased by distinguishing not the bare energy states, but instead the qubit-qubit \emph{eigenstates} that are stationary under the effect of the qubit-qubit coupling and detuning (similarly to the measurement of coupled phase qubits analyzed in \cite{Galiautdinov2012}). Despite the fact that only the main qubit is being measured, the relatively slow measurement process allows the two-qubit system to collapse to these stationary eigenstates, in contrast to what may be naively expected from textbook projective measurements. Notably, these eigenstates have also been shown to be a natural choice for the logical encoding of high-fidelity multi-qubit gates  \cite{Galiautdinov2012} (for similar reasons), which makes multi-qubit eigenstates an unambiguously optimal choice for logical encoding in
realistic parameter regimes.

We further demonstrate for coupled transmon measurement that, in addition to the ensemble-dephasing rate and the qubit-qubit detuning, the measurement fidelity depends on a third important parameter: the readout resonator energy decay rate due to leakage into a transmission line. For decay rates much slower than the qubit-qubit detuning (as is typical in experiments, e.g., \cite{Kelly2015,Corcoles2015,Riste2015}), the leaked resonator field nearly adiabatically follows the qubit-qubit eigenstate to produce little error. However, for decay rates that are comparable to or larger than the qubit-qubit detuning, the resonator decays more rapidly than it can equilibrate with the qubit-qubit eigenstates, causing frustrated dynamics during the measurement. Such rapid resonator decay will primarily couple the leaked field to the bare energy states, while the fast inter-qubit oscillations (compared to the measurement rate) will relate the output signal to the joint qubit-qubit eigenstates. This frustrated dynamics leads to random quantum jumps between the eigenstates. We derive the rate of these quantum jumps (which we call a switching rate) using a model based on fluctuations of the photon number in the resonator, which perturb the two-qubit eigenstates. We show that the switching can be significant for rapid resonator decay, but becomes almost negligible for realistically slow resonator decay. We also derive the measurement error probability resulting from these quantum jumps, and show that it accumulates almost linearly with integration time in an analogous way to the error from energy-decay ($T_1$) processes.

This paper is organized as follows.  In Section~\ref{sec:ensemble}, we introduce the considered system, formulate the problem, and discuss how to model the ensemble-averaged dynamics.  In Section~\ref{sec:eigenbare}, we identify three qualitatively distinct parameter regimes in the ensemble-averaged dynamics: textbook, adiabatic, and frustrated. In Section~\ref{sec:switching}, we study the transition between the adiabatic and frustrated regimes as the resonator decay is varied, by introducing a simple model of a semiclassically fluctuating field in the resonator that produces random quantum jumps between the eigenstates. We derive the average switching rate for these jumps, and numerically confirm this jump behavior by simulating quantum trajectories in the fast resonator decay regime. In Section~\ref{sec:err}, we demonstrate that the contribution of these jumps to the measurement error is nearly linearly increasing with integration time, and find the error minimized over the integration time. We conclude in Section~\ref{sec:conc}.

\section{Considered system and its ensemble-averaged evolution}\label{sec:ensemble}

\begin{figure}[t]
    \begin{center}
        \includegraphics[width=0.7\columnwidth]{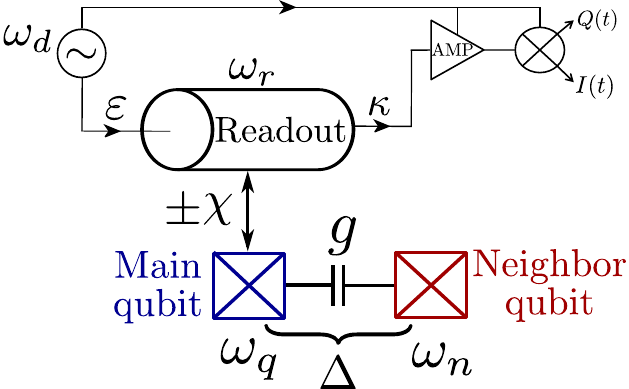}
    \end{center}
  \caption{(Color online) Analyzed system. A measured (main) transmon qubit (blue) with frequency $\omega_{\rm q}$ has capacitive coupling $g$ to a detuned neighboring qubit (red) with a frequency $\omega_{\rm n}$ such that $g \ll |\Delta|$, where $\Delta \equiv \omega_{\rm q} - \omega_{\rm n}$ ($\omega_{\rm q}$ includes the ac Stark shift). A readout resonator with frequency $\omega_{\rm r}$ is dispersively coupled to the main qubit, and thus is frequency-shifted by $\pm \chi$ depending on the main qubit state. During measurement, this resonator is driven with a coherent microwave $\varepsilon$ at a frequency $\omega_{\rm d}$. The field leaks from the resonator (with the energy decay rate $\kappa$) to a transmission line, where it is amplified and mixed with a local oscillator to measure the quadrature $I(t)$ that is sensitive to the qubit-state-dependent phase and amplitude shift. The coupling of the main qubit with the neighboring qubit contributes to the qubit measurement error.
  }
    \label{fig:cQED}
\end{figure}

The archetypal circuit QED system we consider here is shown in Fig.~\ref{fig:cQED}. A superconducting transmon (main qubit) with frequency $\omega_{\rm q}$ (which includes the ac Stark shift) is capacitively coupled to a driven readout resonator with bare frequency $\omega_{\rm r}$, and is also capacitively coupled to another transmon (neighboring qubit) with a detuned bare frequency $\omega_{\rm n}$, such that the qubit-qubit detuning $\Delta\equiv \omega_{\rm q}-\omega_{\rm n}$ is much larger than the qubit-qubit coupling $g$,  $|\Delta|\gg g$. (The role of the neighboring qubit can be played by a bus resonator; we consider a qubit for definiteness.) We assume that the Purcell decay \cite{Purcell1946,Esteve1986,Houck2008} of the main qubit through the resonator has been suppressed by a filter \cite{Jeffrey2014,Reed2010,Sete2015}, and that the resonator and main qubit are sufficiently detuned to treat their coupling as effectively dispersive (implying the rotating wave approximation) \cite{Blais2004,Gambetta2006}. We also assume that the transmon energy levels outside of the qubit subspace are taken into account through renormalization of the state-dependent dispersive shift $\pm\chi$ of the resonator frequency. The readout resonator is additionally driven by a coherent field $\varepsilon$ at a microwave frequency $\omega_{\rm d}$, which then leaks to a transmission line at an energy-decay rate $\kappa$ (the setup can be either in transmission or in reflection). The leaked field is passed through an amplifier and mixed with a local oscillator to perform a homodyne measurement, which isolates the qubit-state-dependent phase and amplitude shift caused by the dispersive coupling; the information-carrying quadrature is denoted as $I(t)$ in Fig.\ \ref{fig:cQED}.

While this measurement procedure is largely understood for a single
qubit coupled to the readout resonator
\cite{Blais2004,Gambetta2006,Gambetta2008,Korotkov2014}, we
investigate here how the addition of the neighboring qubit will
contribute to the  measurement error. Specifically, we wish to find out
whether the wavefunction ``tail'' probability $(g/\Delta)^2$ contributes to the
measurement error or not. In this paper we focus on discriminating the
bare states $\ket{10}$ and $\ket{00}$ (with qubit ordering
convention $\ket{\text{main},\,\text{neighbor}}$) or the states
$\ket{\overline{10}}$ and $\ket{00}$, where $\ket{\overline{10}}$ is
the eigenstate that accounts for the qubit-qubit interaction. We
assume the logic state of zero for the neighboring qubit for simplicity,
without significant loss of generality, because the discrimination
of the states $\ket{11}$ and $\ket{01}$ (or $\ket{11}$ and
$\ket{\overline{01}}$) is a very similar problem. Also, the
discrimination of all four states in the case when both qubits are
measured is a simple generalization of our basic problem. Note that
we do not consider another important question: deterioration of a
superposition $\alpha\ket{00}+\beta\ket{01}$ (or
$\alpha\ket{00}+\beta\ket{\overline{01}}$) after the main qubit
measurement; however, the mechanism of this deterioration is similar
to what we consider. Also note that in an architecture
\cite{Galiautdinov2012,Mariantoni2011}, in which the zero state of
the neighboring qubit is used as a resource to decrease crosstalk,
our assumption of discriminating $\ket{10}$ and $\ket{00}$  (or
$\ket{\overline{10}}$ and $\ket{00}$) is naturally satisfied. We
will refer to the pair of states to be discriminated as
logical 1 and 0.

The logical states are discriminated by integrating the fluctuating
output signal $I(t)$ over time and then comparing the result with a
threshold. Therefore, the discrimination error $P_{\rm err}$
(discussed in more detail in Sec.\ \ref{sec:err}) can be calculated
from the ``overlap'' of the probability distributions of the
integrated result for the two logical states. The error depends on
the chosen threshold (in Sec.\ \ref{sec:err} we will consider the
symmetric and optimal thresholds) and on the integration time. As
will be discussed later, the measurement error $P_{\rm err}$ has a
minimum as a function of the integration time, which is determined
by the rate of ``switching'' (quantum jumps) between the qubit
states, resembling the energy relaxation events.

Our final goal is to find such optimized measurement error for
distinguishing the bare-basis states $\ket{10}$ and $\ket{00}$, and
for distinguishing the eigenbasis states $\ket{\overline{10}}$ and
$\ket{00}$, thus finding which encoding basis is preferable in the
circuit QED measurement. The analysis of a similar question for the
measurement of phase qubits showed \cite{Galiautdinov2012} that
using the  eigenbasis is preferable. In this paper we will obtain a similar
result for the parameter regime of typical
circuit QED measurements
\cite{Kelly2015,Corcoles2015,Riste2015}, even
though the measurement dynamics is significantly more complicated
than for phase qubits. In particular, we will show that in contrast
to what is expected for a textbook projective measurement, the
bare-basis error exceeds $(g/\Delta)^2/2$, while there is no such
limitation for discriminating the eigenstates. For the eigenbasis
the limitation comes from the quantum jumps between the eigenstates
$\ket{\overline{10}}$ and $\ket{\overline{01}}$; however, for
typical experimental parameters this limitation is almost
negligible.

To obtain these results, we first discuss how to model both the coherent and incoherent aspects of the evolution for the ensemble-averaged case. This ensemble-averaged dynamics will be sufficient to identify broad parameter regimes of interest for the coupled-qubit measurement, and to identify which qubit-qubit bases are preserved by the measurement in these regimes, but will be insufficient for understanding and quantifying the measurement error for specific realizations. In Section~\ref{sec:switching}, we will generalize the ensemble-averaged approach to model the individual quantum trajectories, which will allow us to understand and derive the  measurement error induced by the qubit-qubit coupling. Note that we consider only one neighboring qubit, while in practical architectures (e.g., in surface codes) there are several neighbors; however, the generalization of our theory to several neighboring qubits is rather straightforward.

\subsection{Coherent evolution}

The total resonator-qubit-qubit Hamiltonian can be split into five terms,
    \be \label{eq:H}
  H = H_{\rm r} + H_{\rm q} + H_{\rm d} + H_{\rm qr} + H_{\rm qq}.
    \ee
The bare-energy contributions ($\hbar=1$) are
    \be \label{eq:H0}
  H_{\rm r} = \omega_{\rm r} \, a^{\dagger}a,  \quad
  H_{\rm q} = \frac{\omega_{\rm q}^{\rm b}}{2}\sigma_z^{(1)}+\frac{\omega_{\rm n}}{2}\sigma_z^{(2)},
    \ee
where $\sigma_z^{(j)} \equiv \ket{1}\bra{1}_j - \ket{0}\bra{0}_j$ are the Pauli $z$ operators for each qubit ($j=1,2$), $a^\dagger$ ($a$) are the raising (lowering) operators of the coupled resonator mode that satisfy $[a,a^\dagger] =1$, and $\omega_{\rm q}^{\rm b}$, $\omega_{\rm n}$, and $\omega_{\rm r}$ are the bare frequencies of the main qubit, neighboring qubit, and resonator. The resonator drive contribution has the form
\begin{align} \label{eq:Hd}
  H_{\rm d}(t) &= \varepsilon\,e^{-i\omega_{\rm d} t}\,a^\dagger + \varepsilon^*\,e^{i\omega_{\rm d} t}\,a .
\end{align}
The bare energies are modified by the dispersive qubit-resonator coupling
\begin{align}\label{eq:Hqr}
  H_{\rm qr} &= \chi\,\sigma_z^{(1)}\, a^{\dagger}a,
\end{align}
which shifts the resonator frequency by $\pm\chi$ depending on the qubit state or, alternatively, shifts the qubit frequency,
    \be
\omega_{\rm q}=\omega_{\rm q}^{\rm b}+\delta\omega_{\rm q},
    \ee
by the ac Stark shift $\delta \omega_{\rm q}$, depending on the number of photons in the resonator (we include the Lamb shift \cite{Haroche2013,Blais2004,Fragner2008} into $\omega_{\rm q}^{\rm b}$). The qubit-qubit coupling Hamiltonian (assuming the rotating wave approximation) is
\begin{equation}\label{eq:Hqq}
  H_{\rm qq} = g\,(\ket{01}\bra{10} + \ket{10}\bra{01}),
\end{equation}
and we are interested in the case of strongly detuned qubits, $g \ll |\omega_{\rm q} - \omega_{\rm n}|$ (for simplicity we assume $g>0$).

Note that the qubit-qubit coupling in Eq.~\eqref{eq:Hqq} coherently mixes the single-excitation subspace $\{\ket{01},\ket{10}\}$ and produces the eigenstates of the qubit-qubit Hamiltonian $H_{\rm q} + H_{\rm qq}$ that are rotated from the bare states by an angle $\theta = \frac{1}{2}\arctan(2g/\Delta)$,
\begin{equation}\label{eq:eigen}
\begin{aligned}
  \ket{\overline{10}} & =\cos \theta \ket{10} +\sin\theta \ket{01} \approx \sqrt{1-\left(\frac{g}{\Delta}\right)^2}\,\ket{10} + \frac{g}{\Delta}\,\ket{01}, \\
  \ket{\overline{01}} &  =\cos \theta \ket{01} -\sin\theta \ket{10} \approx  \sqrt{1-\left(\frac{g}{\Delta}\right)^2}\,\ket{01}  -\frac{g}{\Delta}\,\ket{10} ,
\end{aligned}
\end{equation}
where
    \be
\Delta\equiv \omega_{\rm q}-\omega_{\rm n}
    \ee
is the (ac Stark-shifted) qubit-qubit detuning and the approximation is to lowest order in $g/|\Delta| \ll 1$. If the measurement process occurs effectively in this eigenbasis, then an initially bare state $\ket{10}$ will collapse into the incorrect eigenstate $\ket{\overline{01}}$ with an error probability $(g/\Delta)^2$, resulting in additional measurement error. In Section~\ref{sec:eigenbare} we will clarify which parameter regimes of the measurement naturally select the eigenstates of Eq.~\eqref{eq:eigen} in this manner. Note that for brevity of notations, in inequalities describing the parameter regimes we will use $\Delta$ instead of $|\Delta|$.

\subsection{Incoherent evolution}

In addition to the coherent evolution given by the Hamiltonian in Eq.~\eqref{eq:H}, the energy in the resonator incoherently decays to a transmission line at the rate $\kappa$. Assuming that all leaked photons may not later return to the resonator, we can model the ensemble-averaged Markovian evolution of the joint qubit-qubit-resonator state with a master equation \cite{Haroche2013}
\begin{equation}\label{eq:master}
  \d\rho = -i[H,\,\rho]\d t + (\kappa\, \d t)\,a\rho a^{\dagger} - \frac{\kappa\, \d t}{2}\,\left(a^{\dagger}a\rho+\rho a^{\dagger}a\right).
\end{equation}
Physically, we can interpret this equation as stating that in a small interval $\d t$ the system does not only coherently evolve with the usual evolution operator
\begin{align*}
  U = \exp(-i H \d t),
\end{align*}
but additionally has one of two distinct incoherent processes happen (e.g., Ref.~\cite{Korotkov2013AppB}):
\begin{enumerate}[(a)]
  \item Each of $N$ photons in the resonator may escape with probability $\kappa\, \d t$, which modifies the resonator state with the decay operator
    \begin{align*}
      M_{\rm decay} &= \sqrt{\kappa\, \d t}\, a.
    \end{align*}
  \item All $N$ photons stay in the resonator with probability $1-\kappa\, \d t\, N$, which modifies the resonator state with the null result (no decay) operator
    \begin{align*}
      M_{\rm null} &= \sqrt{\openone - \kappa\, \d t\, a^\dagger a}.
    \end{align*}
\end{enumerate}
These measurement (Kraus) operators for the incoherent part of the evolution satisfy the usual completeness condition $M^\dagger_{\rm decay}M_{\rm decay} + M^\dagger_{\rm null}M_{\rm null} = \openone$, indicating that the probabilities for each possibility to occur are correctly normalized \cite{bookNielsen2000}. Mixing together both possibilities (i.e., discarding any record of whether the decay happened or not) produces the updated mixed state
\begin{align}\label{eq:masterinc}
  \rho' &= M_{\rm decay}\rho M^\dagger_{\rm decay} + M_{\rm null}\rho M^\dagger_{\rm null}, \\
  &= \kappa\,\d t\,a\rho a^\dagger + \sqrt{\openone - \kappa\, \d t\, a^\dagger a}\,\rho\,\sqrt{\openone - \kappa\, \d t\, a^\dagger a}, \nonumber
\end{align}
that describes the ensemble-averaged evolution for a duration $\d t$. The Hamiltonian evolution is then interleaved between these incoherent updates: $\rho' \mapsto U\rho'U^\dagger$.  Expanding the full increment $\d\rho \equiv U\rho'U^\dagger - \rho$ to linear order in $\d t$ produces the standard master equation form of Eq.~\eqref{eq:master}. We note, however, that the update in Eq.~\eqref{eq:masterinc} is not only conceptually transparent, but (as we checked) is more numerically stable for simulation purposes and in some regimes is faster than solving Eq.~\eqref{eq:master}.

If additional decay channels are present, they can be added phenomenologically to the incoherent sum in Eq.~\eqref{eq:masterinc}.  For example, qubit energy-decay with rate $1/T_1$ and environmental qubit dephasing with rate $\Gamma_{\rm e}$ have the forms
\begin{align*}
  M_{\rm T1} &= \sqrt{\d t/T^{(1)}_1}\,\sigma^{(1)}_- + \sqrt{\d t/T^{(2)}_1}\,\sigma^{(2)}_-, \\
  M_{\rm dephase} &= \sqrt{\Gamma^{(1)}_{\rm e}\, \d t}\,\sigma^{(1)}_z + \sqrt{\Gamma^{(2)}_{\rm e}\, \d t}\,\sigma^{(2)}_z,
\end{align*}
which will modify the null result operator accordingly to include all decay channels $M_k$
\begin{align*}
  M_{\rm null} &= \sqrt{\openone - \textstyle{\sum}_k M_k^\dagger M_k}.
\end{align*}
These additional decay channels correspondingly modify the linear increment in Eq.~\eqref{eq:master} in the standard way.  For simplicity, we will neglect such additional decay channels  in most of what follows, in order to focus solely upon the effects of the neighboring qubit on the measurement fidelity. When we do add these effects, we will assume that $T^{(1)}_1 = T^{(2)}_1 \equiv T_1$ and $\Gamma^{(1)}_{\rm e} = \Gamma^{(2)}_{\rm e} \equiv \Gamma_{\rm e}$.

Now let us briefly review some results for measurement of a {\it single} qubit \cite{Blais2004,Gambetta2006,Gambetta2008,Korotkov2014}, which we will use as a starting point and to introduce notations. For the qubit in the state $|1\rangle$ or $|0\rangle$, the effective frequency of the resonator is $\omega_{\rm r} \pm \chi$ (the upper sign is for the state $|1\rangle$). Then the evolution of the resonator coherent state $\ket{\alpha_\pm}= e^{-|\alpha_\pm|^2/2} \sum_n \alpha_{\pm}^n (n!)^{-1/2}e^{-in\omega_{\rm d} t}|n\rangle$ (we use the rotating frame $e^{-i\omega_{\rm d} t}$) is
    \be
\dot\alpha_{\pm}=-i(\Delta_{\rm r} \pm \chi)\, \alpha_\pm- \frac{\kappa }{2}\,\alpha_\pm -i\varepsilon ,
    \ee
where $\Delta_{\rm r} \equiv \omega_{\rm r} -\omega_{\rm d}$ is the bare resonator-drive detuning. The steady-state solution of this equation is
    \be\label{eq:meana}
  \alpha_{\pm} = \frac{-i\varepsilon}{\kappa/2 + i(\Delta_{\rm r} \pm \chi)},
    \ee
and the corresponding mean photon number is
    \be\label{eq:meann}
  \bar{n}_\pm = |\alpha_\pm|^2 = \bar{n}_{\rm max}\,\frac{\kappa^2}{\kappa^2 + 4(\Delta_{\rm r} \pm \chi)^2},
    \ee
which we expressed via the photon number at exact resonance, $\bar{n}_{\rm max}=4|\varepsilon|^2/\kappa^2$. The ac Stark shift is then \cite{Gambetta2006}
    \be
    \delta \omega_{\rm q} = 2\chi\,\text{Re} (
    \alpha^*_+\alpha_-),
    \ee
and the measurement-induced ensemble dephasing rate is \cite{Gambetta2006}
    \be
  \Gamma_{\rm m} =   2\chi\,\text{Im}(\alpha^*_+\alpha_-) =  \kappa \, \frac{ |\alpha_+ -\alpha_-|^2}{2}.
    \ee
These results can be expressed in terms of $\bar{n}_\pm$ and $\bar{n}_{\rm max}$ as
\begin{align} \label{eq:acstark}
	\delta \omega_{\rm q} &= 2\chi\,\frac{\bar{n}_+\bar{n}_-}{\bar{n}_{\rm max}}\,\left[1 + \frac{4(\Delta_{\rm r}^2 - \chi^2)}{\kappa^2}\right], \\ \label{eq:gammam}
	\Gamma_{\rm m} &= \frac{8 \chi^2}{\kappa}\, \frac{\bar{n}_+\bar{n}_-}{\bar{n}_{\rm max}},
\end{align}
which reduce to the simple formulas \cite{Blais2004,Gambetta2006} $\delta \omega_{\rm q} \approx 2\chi \bar{n}$, $\Gamma_{\rm m} \approx 8 \chi^2\bar{n}/\kappa$ when $\bar{n}_+\approx \bar{n}_-\approx \bar{n}_{\rm max}$.
One of the ways to interpret the measurement-induced dephasing process is as being caused by fluctuations of the ac Stark shift that arise from the fluctuating photon number.
The total ensemble-dephasing rate $\Gamma = \Gamma_{\rm m} + \Gamma_{\rm e}$
 generally includes additional environmental dephasing $\Gamma_{\rm e}$, but we will mostly neglect $\Gamma_{\rm e}$ for simplicity. The measurement-induced ensemble dephasing is related to the \emph{distinguishability time} (sometimes called the ``measurement time'')
\begin{align}\label{eq:tau}
  \tau \equiv (2\eta\Gamma_{\rm m})^{-1},
\end{align}
needed for achieving unit signal-to-noise ratio in the quadrature output, where $\eta\in[0,1]$ is the quantum efficiency of the detection circuit.

We emphasise that these standard results for $\delta \omega_{\rm q}$ and $\Gamma_{\rm m}$ are for the measurement of a single qubit; moreover, they implicitly assume the ``bad cavity limit'' in the sense that the qubit evolution is much slower than $\kappa$ (in this case it is sufficient to consider only coherent states in
the resonator, entangled with the qubit, which leads to
relatively simple formulas). Therefore, we should not expect that these results are directly applicable to our problem, which focuses on evolution involving the neighboring qubit. In particular, when the qubit-qubit detuning $\Delta$ is larger than $\kappa$, the relatively slow fluctuations of the photon number in the resonator will not produce the same dephasing $\Gamma_{\rm m}$ between the states $\ket{10}$ and $\ket{01}$ (as would be expected for infinitely fast fluctuations). Similarly, for $\Delta \gg \kappa$ the ac Stark shift contribution to $\Delta$ is supposed to be governed mainly by $\bar{n}_+$ (or $\bar{n}_-$ if the main state is $\ket{01}$) rather than given by Eq.\ (\ref{eq:acstark}). Even though $\Gamma_{\rm m}$ in Eq.\ (\ref{eq:gammam}) does not in general describe the ensemble dephasing between $\ket{10}$ and $\ket{01}$, in this paper we will extensively use $\Gamma_{\rm m}$ defined in Eq.\ (\ref{eq:gammam}) as a notation.

\section{Eigenstates vs. bare states}\label{sec:eigenbare}

A master equation is incapable of describing the fidelity of the qubit measurement, even in principle, so we will be forced to consider the individual quantum trajectories in Section~\ref{sec:switching}. Nevertheless, even without a more detailed trajectory description we can already answer the most basic question about the qubit measurement: does the ensemble-averaged evolution faithfully preserve a logical qubit basis?

\begin{figure}[t]
    \begin{center}
        \includegraphics[width=\columnwidth]{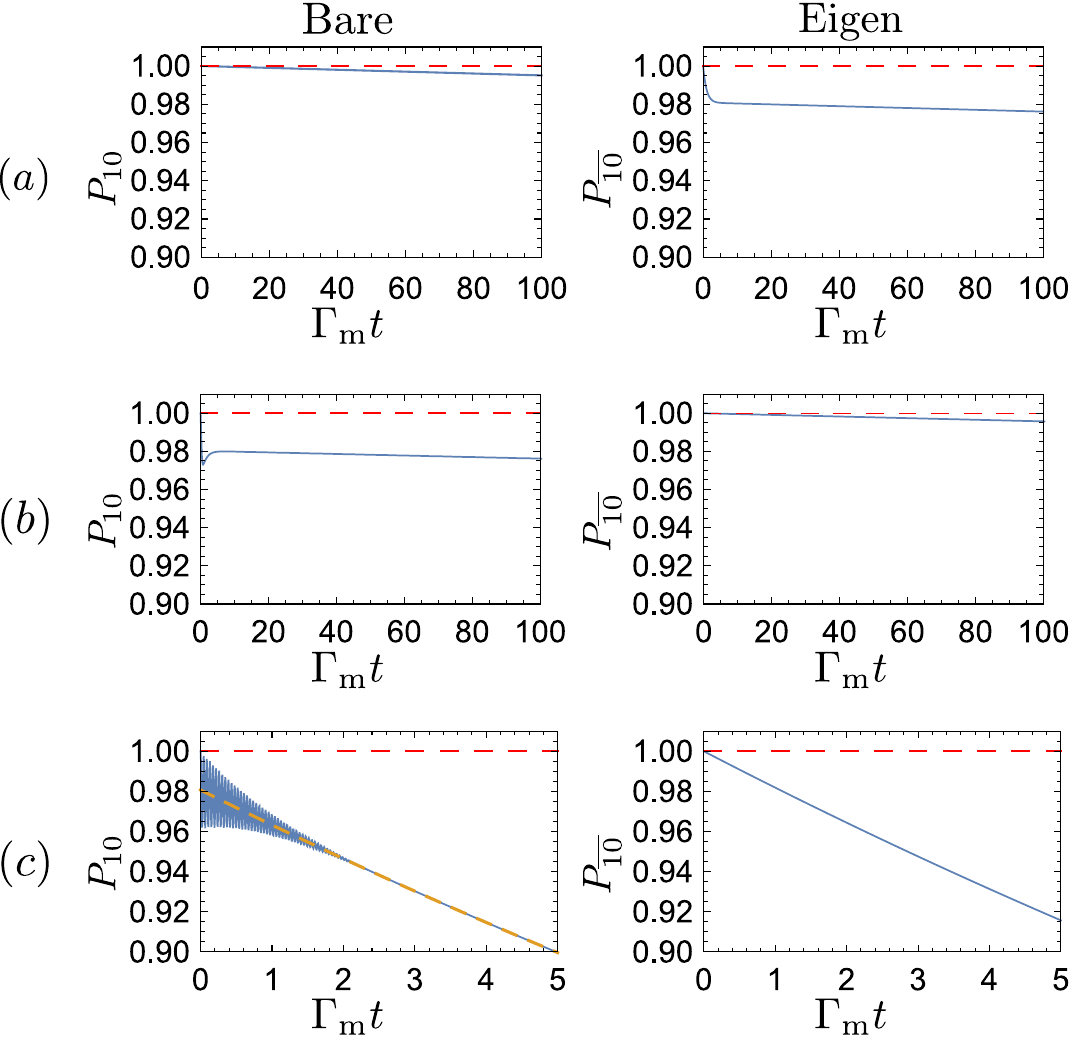}
    \end{center}
    \caption{(Color online) Blue lines: ensemble-averaged evolution of the population $P_{10}$ of the bare-basis state $|10\rangle$ when the evolution starts in this state (left panels) and the population $P_{\overline{10}}$ of the eigenstate $|\overline{10}\rangle$ when starting in this state (right panels). The dashed red lines show the initial value 1 of all blue lines for reference. The time is normalized by the ensemble-dephasing rate $\Gamma_{\rm m}$ due to measurement; we assume fixed qubit-qubit coupling and detuning with $g/\Delta = 1/10$ for all regimes.
 (a) Textbook regime with $\Delta \ll \Gamma_{\rm m} \ll \kappa$, using directly applied qubit-dephasing of $\Gamma_{\rm m}/\Delta = 20$ for simplicity (i.e., assuming $\kappa \to \infty$). The bare state $\ket{10}$ is best preserved by the evolution, but slowly decays at the rate $2g^2/\Gamma_{\rm m}$, while the eigenstate population $P_{\overline{10}}$ additionally drops by approximately $2(g/\Delta)^2$.
 (b) Adiabatic (experimental) regime with $(\Gamma_{\rm m}, \kappa) \ll \Delta$, using $\kappa/\Delta = 10^{-1}$ and $\Gamma_{\rm m}/\Delta = 10^{-2}$, set by assuming a weak response $\chi/\kappa = 3.5\times10^{-2}$ and a resonator drive $\omega_{\rm d}=\omega_{\rm r}$ with power tuned to produce the steady-state photon number $\bar{n} = 10$. The eigenstate $\ket{\overline{10}}$ is best preserved by the evolution, but slowly decays (analogously to the textbook regime for $P_{10}$), while the bare population $P_{10}$ additionally drops by $2(g/\Delta)^2$.
 (c) Frustrated regime with $\Gamma_{\rm m} \ll \Delta \ll \kappa$, using $\kappa/\Delta = 10$ and $\Gamma_{\rm m}/\Delta = 10^{-4}$, keeping the same $\chi/\Delta$ and $\bar{n}$ as in the adiabatic regime. The bare state population $P_{10}$ drops by $2(g/\Delta)^2$ compared to the eigenstate population $P_{\overline{10}}$, and both populations show rapid decay.
 The decay rate seen in regimes (b) and (c) matches the analytical results for averaged incoherent quantum jumps between the eigenstates (see Fig.~\ref{fig:switching}), an example of which is shown here in the bare (c) plot as the overlaid dashed yellow curve.}
    \label{fig:decay}
\end{figure}

To answer this question, we simulate the full master equation in Eq.~\eqref{eq:master} [equivalently, Eq.~\eqref{eq:masterinc} can be iterated] starting in either a bare state $\ket{10}$, or an eigenstate $\ket{\overline{10}}$, with the resonator in an initial ground state for simplicity (the simulation starting in the state $\ket{00}$ is trivial). When starting in $\ket{10}$, we calculate the evolution of the bare state population $P_{10}$, and when starting in $\ket{\overline{10}}$, we calculate the eigenstate population $P_{\overline{10}}$ (see the left and right panels in Fig.\ \ref{fig:decay}). If one of these populations remains very close to 1, then we infer that the corresponding basis is faithfully preserved by the measurement dynamics.

As shown in Fig.~\ref{fig:decay}, from these simulations we identify three parameter regimes that have qualitatively different behaviors (using $g\ll \Delta$ and $\Gamma_{\rm e} =0$):
\begin{enumerate}[(a)]
  \item $\Delta  \ll (\Gamma_{\rm m}, \kappa)$ : textbook --- \emph{almost stable bare state},
  \item $(\Gamma_{\rm m}, \kappa) \ll \Delta $ : adiabatic --- \emph{almost stable eigenstate},
  \item $\Gamma_{\rm m} \ll \Delta  \ll \kappa$ : frustrated --- \emph{unstable eigenstate}.
\end{enumerate}
The parameters used for each of these regimes are detailed in the
caption for Fig.~\ref{fig:decay}. As expected from the similar
analysis for measurement of phase qubits \cite{Galiautdinov2012},
for $\Gamma_{\rm m} \ll \Delta$ the measurement effectively occurs
in the eigenbasis, while the traditional (textbook) bare-basis
measurement requires $\Gamma_{\rm m} \gg \Delta$. However, transmon
qubits have an additional important parameter that has no analogue
in phase qubits: the resonator energy-decay rate $\kappa$. As we
will see, the relative magnitudes of $\kappa$ and $\Delta$ determine
the ``stability'' of the eigenbasis.

In the regime (a), the resonator empties and the system dephases
much faster than the qubit-qubit evolution, so the \emph{bare
states} $\ket{10}$ and $\ket{00}$ are preserved as the optimal
logical basis, just as we would expect from a textbook projective
measurement. That is, our numerical simulation in
Fig.~\ref{fig:decay}(a) shows that the bare state population
$P_{10}$ is preserved practically at $1$, while the eigenstate
population $P_{\overline{10}}$ (when starting with
$\ket{\overline{10}}$) drops by roughly $2(g/\Delta)^2$ during the
transient (collapse) evolution. [Here one factor of $(g/\Delta)^2$
stems from the physical collapse of the eigenstate to an incoherent
mixture of the single-excitation bare states $\ket{10}$ and
$\ket{01}$, while the second factor $(g/\Delta)^2$ comes from
plotting the eigenstate population.] At a much longer time scale the
bare-basis population gradually decreases because non-zero $g$ makes
the measurement not fully projective, leading to rare transitions
(jumps) between the states $\ket{10}$ and $\ket{01}$. Note that for numerical simplicity in the regime (a) we simulated the evolution assuming $\kappa\gg \Gamma_{\rm m}$, so that the qubits and resonator remain effectively disentangled (qubit entanglement with the emitted field is not important for the master equation approach). With this approximation, we can simplify Eq.~\eqref{eq:master} by reducing it to a two-qubit Hilbert space and taking into account the interaction with the resonator by applying the dephasing with the rate $\Gamma_{\rm m}$ to the measured qubit. In this case the transient evolution
occurs on the time scale $\Gamma_{\rm m}^{-1}$ and the population $P_{10}$
decays with the rate $2g^2/\Gamma_{\rm m}$. The textbook regime (a) is
most easy to understand and analyze. However, we emphasize that this
regime is not realized in realistic experiments with transmons, in
which typically $\Delta \gg \Gamma_{\rm m}$.

In the adiabatic regime (b), which more closely describes recent
experiments \cite{Kelly2015,Corcoles2015,Riste2015}, the resonator empties and the system dephases more slowly
than the qubit-qubit evolution, so the \emph{eigenstates}
$\ket{\overline{10}}$ and $\ket{00}$ are preserved as the optimal
logical basis. That is, our numerical simulation in
Fig.~\ref{fig:decay}(b) shows that the eigenstate population
$P_{\overline{10}}$ is preserved at almost $1$, in contrast to the
textbook regime, while the bare state population $P_{10}$ (when
initially 1) drops by roughly $2(g/\Delta)^2$ within the collapse
timescale (this timescale is $\Gamma_{\rm m}^{-1}$ if $\kappa\gg \Gamma_{\rm m}$, while for $\kappa\ll \Gamma_{\rm m}$ everything is determined by transients). Again, in this drop one
factor of $(g/\Delta)^2$ comes from the collapse into an incoherent
mixture of single-excitation eigenstates $\ket{\overline{10}}$ and
$\ket{\overline{01}}$, while the second factor comes from plotting
the bare state population. At longer time scales, we also observe in
Fig.~\ref{fig:decay}(b) that the eigenstate population
$P_{\overline{10}}$ decays exponentially at a very slow rate. This
occurs because of rare transitions (jumps) between the eigenstates
$\ket{\overline{10}}$ and $\ket{\overline{01}}$, discussed in more
detail later.

The frustrated regime (c) differs from the adiabatic
regime (b) only by the relative magnitude of the resonator decay
$\kappa$ and the qubit-qubit detuning $\Delta$, $\kappa \gg\Delta$.
Nevertheless, this regime dramatically amplifies the exponential
decay process observed at long times in the adiabatic regime (b).
The rapid decay seen in Fig.~\ref{fig:decay}(c) occurs for both bare and
eigenstates, so that neither of these bases is good for preserving a logical state. This occurs because fast oscillations $\Delta$ (compared to dephasing $\Gamma_{\rm m}$) favor the eigenbasis, while even faster decay $\kappa$ makes the outgoing photons sensitive to the bare basis.

\begin{figure}[t]
    \begin{center}
\includegraphics[width=0.9\columnwidth]{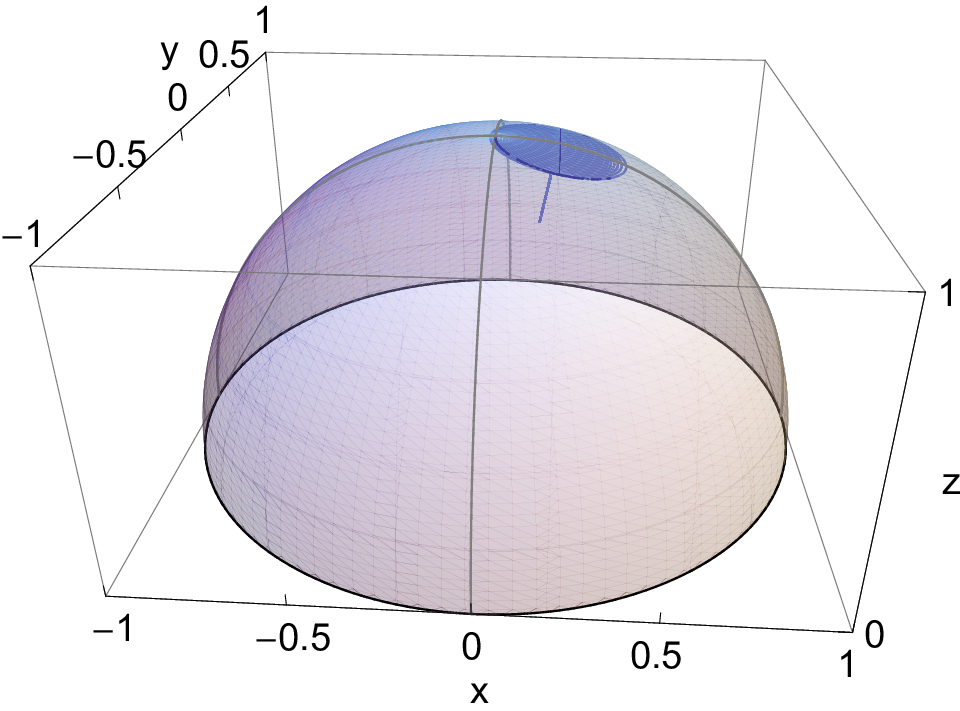} \\
    \vspace{1em}
\includegraphics[width=0.7\columnwidth]{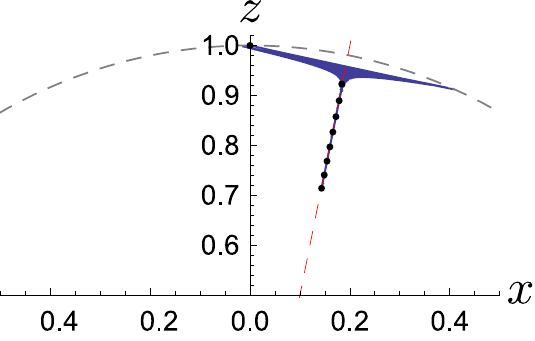}
    \end{center}
\caption{(Color online) Qubit-qubit single-excitation ensemble-averaged evolution, using
bare Bloch sphere coordinates defined by $x = \ket{10}\bra{01} +
\ket{01}\bra{10}$, $y = -i(\ket{10}\bra{01} - \ket{01}\bra{10})$ and
$z = \ket{10}\bra{10} - \ket{01}\bra{01}$, and parameters $g/\Delta
= 1/10$, $\kappa/\Delta = 1$, and $\Gamma_{\rm m} /\Delta = 1/100$
(with $\bar{n}=2$ and correspondingly $\chi/\kappa=1/40$). An initially bare state
$\ket{10} \equiv (0,0,1)$ oscillates rapidly around the tilted axis
corresponding to the eigenbasis
$\{\ket{\overline{10}},\ket{\overline{01}}\}$, and approaches this
axis, indicating the gradual collapse, which produces an incoherent
mixture of the eigenstates. At longer times, the ensemble averaged
state continues moving along the eigenstate axis at a slow rate,
indicating an additional classical mixing process. (top) 3D plot of
$(x,y,z)$ evolution, showing the spiraling evolution to the
eigenstate axis and then along the axis, simulated for $\Gamma_{\rm
m} t \in [0,40]$. (bottom) Slice of x-z plane, with
sphere surface shown as the dashed gray curve and the eigenstate
axis shown as the dashed red line tilted from the bare $z$ axis by
the angle $2\theta = \arctan(2g/\Delta)$. Black dots show time intervals of
$\Gamma_{\rm m} t = 5$.}
  \label{fig:collapse}
\end{figure}
In both regimes (b) and (c), the system collapses to the \emph{eigenstates}, after which the state may jump between the eigenstates. This behavior is evidenced in
Fig.~\ref{fig:collapse}, showing the ensemble-averaged evolution in the Bloch sphere representation of the qubit-qubit single-excitation subspace. The ratio $\kappa/\Delta =1$ is chosen in between the regimes (b) and (c). The initially bare state rapidly oscillates around the eigenstate axis as it spirals into
this axis on average, indicating that the initially bare state
collapses to an incoherent mixture of the eigenstates. After that the exponential decay occurs along the eigenstate axis of the Bloch sphere, indicating that it arises solely from a \emph{classical mixing} process that scrambles those eigenstates.

The physical origin of the exponential decay seen in regimes (b) and (c) is not apparent from examining the ensemble-averaged behavior of the master equation alone, but we shall see that this decay can be interpreted as arising from averaging random quantum jumps between the eigenstates that occur during the continuous measurement process. For the remainder of this
paper we will mostly focus on understanding the transition between the
adiabatic regime (b) and the frustrated regime (c) as $\kappa$ is
varied with respect to $\Delta$.

We also briefly note that in principle there is a fourth parameter regime: (d) $\kappa \ll \Delta \ll \Gamma_{\rm m}$. We do not consider this regime here, since in this case it is difficult to clearly pose the problem of finding a preferable measurement basis without focusing solely on the ring up evolution for the resonator, and since this regime is not relevant to actual experiments.

\section{Quantum jumps in eigenbasis}\label{sec:switching}

In this section we focus on understanding the exponential decay at
long times in Figs.~\ref{fig:decay} and \ref{fig:collapse} for the
adiabatic and frustrated regimes (b) and (c). The ensemble-averaged
simulation suggests that after an initial state collapses to one of
the two eigenstates $\{\ket{\overline{10}},\ket{\overline{01}}\}$,
these eigenstates then become further mixed at a rate that depends
on the relative magnitude of the resonator decay $\kappa$ and the
qubit-qubit detuning $\Delta$. As we will soon see using quantum
trajectory simulations, this mixing process can be identified as
stochastic quantum jumps between otherwise stabilized eigenstates.

Treating these jumps semiclassically as telegraph noise, we surmise
there must exist two unidirectional \emph{switching rates}
$\Gamma^{\pm}_{\text{sw}}$ for randomly transitioning from the state
$\ket{\overline{10}}$ to $\ket{\overline{01}}$ ($-$) or vice versa
($+$). The eigenstate population therefore should obey the simple
ensemble-averaged rate equation
  \begin{align}\label{eq:telegraph}
  \dot{P}_{\overline{10}} &=
   - \Gamma^{-}_{\text{sw}}\,P_{\overline{10}}
  +\Gamma^{+}_{\text{sw}}\,(1 - P_{\overline{10}})
 ,
  \end{align}
where we used $P_{\overline{10}}+P_{\overline{01}}=1$. In
particular, the solution of this equation starting with
$P_{\overline{10}}(0) = 1$ is
\begin{align}\label{eq:popdrop}
  P_{\overline{10}}(t) &=
   \frac{\Gamma^{+}_{\text{sw}}}{\Gamma^{+}_{\text{sw}}+
\Gamma^{-}_{\text{sw}}}
 + \frac{\Gamma^{-}_{\text{sw}}}{\Gamma^{+}_{\text{sw}} +
\Gamma^{-}_{\text{sw}}}\, e^{-(\Gamma^{+}_{\text{sw}} +
\Gamma^{-}_{\text{sw}})t},
\end{align}
it eventually saturates at the population
$\Gamma^{+}_{\text{sw}}/(\Gamma^{+}_{\text{sw}} +
\Gamma^{-}_{\text{sw}})$, and has an initial decay slope of
$\Gamma^{-}_{\text{sw}}$. If the switching rates are equal, the
solution will eventually reach the maximally mixed eigenpopulation
of $1/2$ (i.e., the center of the Bloch sphere in
Fig.~\ref{fig:collapse}).
We derive the switching rates for this model in the next
section, after which we will describe how to simulate the quantum
trajectories that show this switching behavior explicitly.

\subsection{Switching rate}\label{sec:switching-rate}

In order to calculate the rates $\Gamma_{\rm sw}^\pm$ of jumps
between the eigenstates $\ket{\overline{10}}$ and
$\ket{\overline{01}}$ in the slow dephasing regime $\Gamma_{\rm m} \ll
\Delta$ of Fig.~\ref{fig:decay}(b,c), we will take literally the
interpretation of the ensemble dephasing $\Gamma_{\rm m}$ in
Eq.~\eqref{eq:gammam} as being due to a fluctuating number of photons
in the resonator, causing a fluctuating ac Stark shift. Moreover,
we will treat the fluctuating photon numbers $n_\pm(t) = \bar{n}_\pm
+ \delta n_\pm(t)$ as classical variables, with the means
$\bar{n}_\pm$ given in Eq.~\eqref{eq:meann} and fluctuations $\delta
n_\pm(t)$ having temporal correlations \cite{Blais2004,Clerk2010}
\begin{align}\label{eq:poisson}
  \langle \delta n_\pm(t) \, \delta n_\pm(0) \rangle = \bar{n}_\pm\,e^{-\kappa |t|/2}.
\end{align}
Here the upper sign corresponds to the state $\ket{\overline{10}}$,
for which the main qubit is practically in the state $\ket{1}$,
while the lower sign is for $\ket{\overline{01}}$. We will be mostly
interested in the switching rate $\Gamma_{\rm sw}^-$ for the process
$\ket{\overline{10}}\rightarrow \ket{\overline{01}}$, which is
caused by fluctuations of $n_+(t)$; however, for completeness we
calculate both switching rates (the upper sign in all equations
below is sufficient to find $\Gamma_{\rm sw}^-$). Note that the
decay rate of $\kappa/2$ in Eq.\ (\ref{eq:poisson}) is consistent
with the decay of classical energy fluctuations in a pumped
resonator (in contrast to the energy decay $\kappa$ in an unpumped
resonator). Also note that here we neglected the oscillations of the correlator with frequency $\Delta_{\rm r}\pm \chi$ (discussed later).

The fluctuating number of photons $n_\pm (t)$ causes a fluctuating
ac Stark shift $2\chi n_\pm(t)$ [as follows from the dispersive
coupling of Eq.~\eqref{eq:Hqr}], which leads to a fluctuating
qubit-qubit detuning $\Delta+\delta \Delta$, with $\delta \Delta (t)
= 2\chi\,\delta n_\pm(t)$. This in turn produces a fluctuating
effective coupling $\tilde{g}(t)$ between the stationary eigenstates
$\ket{\overline{10}}$ and $\ket{\overline{01}}$, since they are no
longer true eigenstates for the detuning $\Delta+\delta \Delta$. The
fluctuations $\delta n_+(t)$ for the state $\ket{\overline{10}}$
produce the coupling
\begin{equation}\label{eq:dgq}
  \tilde{g}_+ (t) =\bra{\overline{01}}\delta H\ket{\overline{10}}=
  -\frac{g}{\Omega}\, \delta\Delta (t)
  = -2\frac{g}{\Omega}\,\chi\, \delta n_+ (t),
\end{equation}
while for the state $\ket{\overline{01}}$ the fluctuations $\delta
n_-(t)$ are somewhat different, producing
    \begin{equation}\label{eq:dgq-2}
  \tilde{g}_- (t) =\bra{\overline{10}}\delta H\ket{\overline{01}}=
 -2\frac{g}{\Omega}\,\chi\, \delta n_- (t),
  \end{equation}
where in the single-excitation subspace $\delta H= (\delta\Delta/2)
\, (\ket{10}\bra{10}-\ket{01}\bra{01})$ and
    \be
\Omega = \Delta\sqrt{1 + (2g/\Delta)^2} \approx \Delta
    \ee
is energy difference between $\ket{\overline{10}}$ and
$\ket{\overline{01}}$ (we omit the subscripts in $\Delta_\pm$,
$\Omega_\pm$ and $\delta \Delta_\pm$ for brevity). The derivation of
Eq.\ (\ref{eq:dgq}) is very simple when $g\ll \Delta$. Then the true
eigenstates should correspond to the rotation angle $\theta\approx
g/(\Delta +\delta \Delta)$ from the bare basis instead of the angle
$\theta\approx g/\Delta$ for $\ket{\overline{10}}$ and
$\ket{\overline{01}}$. The additional angle,
$\delta\theta\approx -g\,\delta\Delta/\Delta^2$, is the rotation $\tilde g/\Delta$
between the true and stationary eigenbases. Thus we obtain $\tilde g
=-(g/\Delta) \, \delta\Delta$, which is Eq.\ (\ref{eq:dgq}) with
$\Omega\approx \Delta$. In the exact derivation we can use
$\theta={\rm arctan}(2g/\Delta)/2$, then the derivative is
$d\theta/d\Delta=-g/\Omega^2$, which should be equal to $(\tilde
g/\Omega )/\delta\Delta$; this gives Eq.\ (\ref{eq:dgq}).

The fluctuating effective coupling $\tilde g$ between the
eigenstates $\ket{\overline{10}}$ and $\ket{\overline{01}}$ leads to
a gradual mixing between them, which corresponds to random jumps
between the eigenstates in the approach of quantum trajectories. We
can find the rate $\Gamma_{\rm sw}$ of these jumps by starting with
one of the eigenstates and equating $\Gamma_{\rm sw} t$ with the
population of the other eigenstate, which follows from the solution
of the Schr\"odinger equation with the coupling  $\tilde g$. Thus,
to lowest order in $\tilde{g}$ we find the switching rate
  \begin{align}
  \Gamma^{\mp}_{\text{sw}} = \left< \frac{1}{t}\left| \int_0^{t}
  \tilde{g}_\pm(t')\, e^{\pm i\Omega (t-t')}\,\mathrm{d}t' \right|^2 \right> ,
  \label{eq:Gamma-sw-1}\end{align}
where the brackets mean averaging over the random realizations of
$\tilde g(t)$. This equation formally depends on time $t$; however,
there is actually no time dependence for sufficiently long $t$, for
which the evolution can be physically described by a switching rate.
This can be seen by expressing the square of the windowed Fourier
transform in Eq.\ (\ref{eq:Gamma-sw-1}) via the (two-sided) spectral
density $S_{\tilde g_\pm}(\omega)$ of $\tilde{g}_\pm(t)$:
$\Gamma^{\mp}_{\text{sw}} =\int_{-\infty}^\infty S_{\tilde g_\pm}
(\pm\Omega+\omega) [1-\cos(\omega t)](\pi t \omega^2)^{-1}d\omega$.
Therefore, at sufficiently long times $\Gamma^{\mp}_{\text{sw}} =
S_{\tilde g_\pm} (\pm \Omega)$, which does not depend on time.
Because of the linear relations (\ref{eq:dgq}) and (\ref{eq:dgq-2})  between $\tilde g_\pm
(t)$ and $\delta n_\pm (t)$, their spectral densities are related as
$S_{\tilde g_\pm} (\Omega)= (2\chi g/\Omega)^2 S_{\delta n_\pm}
(\Omega)$, therefore
    \be
  \Gamma^{\mp}_{\text{sw}} = (2\chi g/\Omega)^2 S_{\delta n_\pm}
  (\pm \Omega).
    \label{eq:Gamma-sw-Sn}\ee
[Note that for classical fluctuations $\delta n_\pm(t)$ the spectral
density is symmetric, $S_{\delta n_\pm} (-\Omega) = S_{\delta n_\pm}
(\Omega)$; however, we keep the sign of $\Omega$ in Eq.\
(\ref{eq:Gamma-sw-Sn}) to discuss the asymmetric case later.] The
(two-sided) spectral density $S_{\delta n_\pm}$ can be found via the
Wiener-Khinchin theorem \cite{bookStoica1997} $ S_{\delta
n_\pm}(\Omega )=\int_{-\infty}^{\infty} \mean{\delta n_\pm(t) \,
\delta n_\pm(0)}\, e^{-i\Omega t}\,\mathrm{d}t$, so that using Eq.\
(\ref{eq:poisson}) we obtain the switching rate
    \be
     \Gamma^{\mp}_{\text{sw}} = \frac{2g^2}{\Omega^2}\,
     \frac{8\chi^2\bar{n}_\pm}{\kappa} \,
     \frac{\kappa^2}{\kappa^2+4\Omega^2}.
    \label{eq:Gamma-sw-cl}\ee
This result obviously assumes $\Gamma^{\mp}_{\text{sw}} \ll |\Omega |$ and is not applicable during the initial transient evolution due to collapse.

Note that the term $8\chi^2\bar{n}_\pm /\kappa$ in Eq.\ (\ref{eq:Gamma-sw-cl}) is
similar to the measurement-induced dephasing $\Gamma_{\rm m}$ given by
Eq.\ (\ref{eq:gammam}), but it depends on $\bar{n}_+$ for
$\Gamma^{-}_{\text{sw}}$ (or on $\bar{n}_-$ for
$\Gamma^{+}_{\text{sw}}$) rather than the combination
$\bar{n}_+\bar{n}_-/\bar{n}_{\rm max}$ in Eq.\ (\ref{eq:gammam}). In
the case when $\bar{n}_+\approx \bar{n}_-\approx \bar{n}_{\rm max}$
(which occurs when $|\Delta_{\rm r}\pm\chi|\ll \kappa$) we obtain $\Gamma^{+}_{\text{sw}}\approx
\Gamma^{-}_{\text{sw}}\equiv \Gamma_{\text{sw}}$ with
    \be
 \Gamma_{\text{sw}} \approx 2\Gamma_{\rm m}\,\frac{g^2}{\Delta^2}\,
 \frac{\kappa^2}{\kappa^2+4\Delta^2},
    \label{eq:Gamma-sw-simple}\ee
where we also used $\Omega\approx \Delta$ since $g\ll \Delta$. Note
that in the regime $\kappa\gg\Delta$ [as in Fig.\
\ref{fig:decay}(c)] the last factor in Eq.\
(\ref{eq:Gamma-sw-simple}) is close to 1, and the switching rate is
rather large,  $\Gamma_{\text{sw}} \approx 2\Gamma_{\rm m}(g/\Delta)^2$,
while in the regime $\kappa\ll\Delta$ [as in Fig.\
\ref{fig:decay}(b)] the switching rate is additionally suppressed by
the factor $(\kappa /2\Delta)^2$.

\begin{figure}[t]
  \includegraphics[width=0.97\columnwidth]{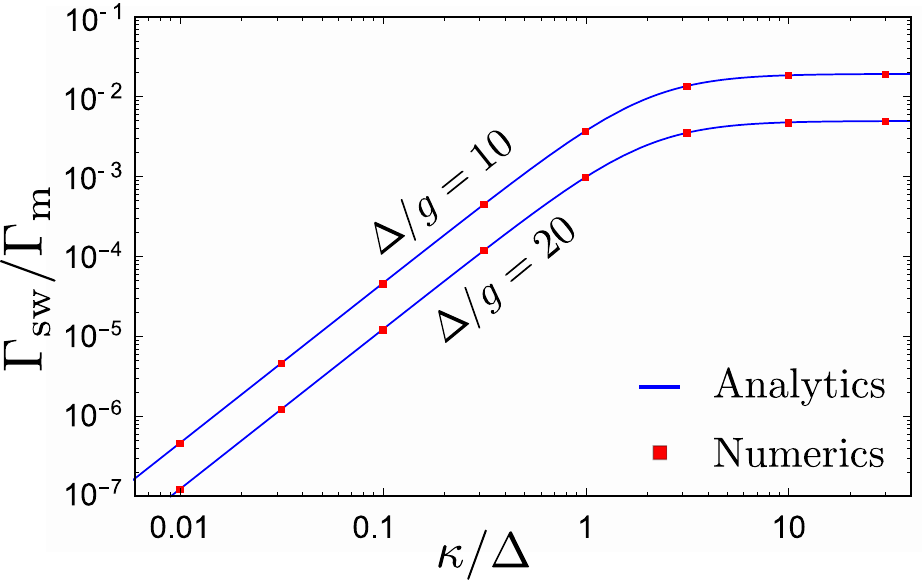}
   \caption{(Color online) Switching rate vs. cavity decay rate. Blue solid line: switching rate analytics $\Gamma_{\text{sw}}/\Gamma_{\rm m} = 2(g/\Omega)^2/[1 +
(2\Omega/\kappa)^2]$ as a function of the ratio $\kappa/\Delta$. Red
boxes: numerical switching rate obtained from solving the master
equation and extracting the decay rate $\Gamma_{\text{sw}}$ from
fitting $P_{\overline{10}}(t)$ [as shown in the right panels of
Figs.~\ref{fig:decay}(b,c)] with Eq.\ (\ref{eq:popdrop}), assuming
$\Gamma_{\text{sw}}^+=\Gamma_{\text{sw}}^-$. A typical relative
difference between the numerical and analytical results is about
$10^{-3}$, which is comparable to an inaccuracy from the fitting
procedure. The used parameters are: $g/\Delta =
1/10$ and $1/20$, $\bar{n}=10$, $\Delta_{\rm r}=0$, and $\chi/\Delta =10^{-3}$. }
  \label{fig:switching}
\end{figure}
We have numerically verified Eq.\ (\ref{eq:Gamma-sw-cl}) for the
switching rates by comparing the telegraph noise solution of
Eq.~\eqref{eq:popdrop} to the ensemble-averaged population decay
obtained from simulating the full master equation \eqref{eq:master} for a
range of $\kappa/\Delta$. The comparison is plotted in
Fig.~\ref{fig:switching}, showing excellent agreement. A typical
mismatch between the analytics and numerics is on the order of
0.1\%, which is comparable to the inaccuracy from the numerical
fitting procedure. Most importantly, in Fig.\
\ref{fig:switching} we see a strong suppression of the switching
rate at $\kappa/\Delta \ll 1$.

Solution of Eq.\ (\ref{eq:telegraph}) with the switching rates given by Eq.\ (\ref{eq:Gamma-sw-cl}) is sufficient to describe the ensemble-averaged evolution when the initial state is an eigenstate. If this is not the case, we need to include collapse of the initial state into the eigenbasis. In particular, for the bare initial state $\ket{10}$, the evolution in Eq.~(\ref{eq:telegraph}) effectively starts with $P_{\overline{10}}(0)=\cos^2 \theta$. As an example, the yellow line in the left panel of Fig.\ \ref{fig:decay}(c) shows such evolution, which is then converted back into the bare basis. While this simple approach does not describe the transient (collapse) dynamics, it accurately describes the evolution after that.

Our derivation for the switching rates $\Gamma^\mp_{\text{sw}}$ in this section has
been based on treating fluctuations $\delta n_\pm (t)$ as classical
fluctuations. The quantum nature of these fluctuations leads to an
asymmetric spectral density \cite{Clerk2010} $S_{\delta
n\pm}(\Omega)= \bar{n}_\pm \kappa/[(\kappa/2)^2+(\Omega -\Delta_{\rm r}
\mp \chi)^2]$. Inserting this formula into Eq.\
(\ref{eq:Gamma-sw-Sn}), we obtain
$ \Gamma^\mp_{\text{sw}} = 2 (g/\Omega)^2
 (8\chi^2\bar{n}_\pm /\kappa)
 \kappa^2/[\kappa^2+4(\pm\Omega -\Delta_{\rm r}
\mp \chi)^2],$
which introduces a slight correction compared to Eq.\
(\ref{eq:Gamma-sw-cl}). Physically, this formula says that if the extra photon energy $-\Delta_{\rm r}\mp\chi$ is positive, this helps the switching process with increase of energy, and vice versa. Even though this correction is very minor in
the typical case, our numerical results using the master equation
confirm the presence of this correction. However, our numerical
results are more consistent with the combination $\kappa^2+4 (\pm\Omega
-\Delta_{\rm r} \pm \chi)^2$ in the denominator of the equation. This combination means that the process depends on the extra photon energy $-\Delta_{\rm r} \pm \chi$ after the switching (which changes the resonator frequency by $\pm 2\chi$) instead of the extra energy $-\Delta_{\rm r} \mp \chi$ before the switching. Note that the logic of our derivation cannot correctly take into account the change of the resonator frequency during switching. Also, numerical results for some parameters are not consistent with the combination $-\Delta_{\rm r} \pm \chi$ as well (indicating a possible presence of a parameter-dependent coefficient in front of $\pm \chi$). Therefore, we are confident only in the correction of Eq.\ (\ref{eq:Gamma-sw-cl}) due to $\Delta_{\rm r}$,
\begin{align}\label{eq:switchingrate}
 \Gamma^\mp_{\text{sw}} &= \frac{2g^2}{\Omega^2}\,
 \frac{8\chi^2\bar{n}_\pm}{\kappa}\,
 \frac{\kappa^2}{\kappa^2+4(\pm\Omega -\Delta_{\rm r})^2},
\end{align}
omiting the dependence on $\chi$ in the denominator.

Note that if additional environmental dephasing $\Gamma_{\rm e}$ is
included in the master equation, it will contribute a similar term
of $(2\Gamma_{\rm e}) (g/\Omega)^2$ to both the up and down switching
rates equally. Environmental energy-decay will also effectively
contribute a term $(1/T_1)$ to only the down switching rate
$\Gamma^{-}_{\text{sw}}$. However, while such energy decay may be
qualitatively similar in its effect on the excited population
$P_{\overline{10}}$, it is intrinsically different from the
eigenstate switching behavior derived here since it transfers the
excitation to the ground state $\ket{00}$ outside the
single-excitation subspace, instead of switching to
$\ket{\overline{01}}$.

We also note that the mechanism discussed here of switching between
the eigenstates $\ket{\overline{10}}$ and $\ket{\overline{01}}$ is physically
similar to the mechanism of ``dressed dephasing''
\cite{Boissonneault2008,Boissonneault2009}, in
which the role of the two-qubit coupling is played by the
Jaynes-Cummings coupling between the qubit and resonator.

\subsection{Quantum trajectory simulations}\label{sec:trajs}

In order to justify our understanding of the exponential decay in
Fig.~\ref{fig:decay}(b,c) as resulting from quantum jumps, we must
go beyond the master equation in Eq.~\eqref{eq:master} and consider
more detailed \emph{quantum trajectories}
\cite{Carmichael1993book,Wiseman1993,Korotkov1999,Gambetta2008,Wiseman2010} (which have been confirmed experimentally with superconducting qubits \cite{Murch2013,Hatridge2013,De_Lange2014,Roch2014}).
In this approach we simulate individual realizations of the
evolution due to measurement, rather than ensemble-averaged
dynamics. In particular, in this case there is no
measurement-induced {\it dephasing} (i.e., a change of the qubit phase);
instead, the gradually-acquired information obtained from
measurement causes continuous stochastic ``attraction'' to the
states $\ket{0}$ and $\ket{1}$ of the measured qubit (random motion
along the meridians on the Bloch sphere). After ensemble averaging,
these two evolutions produce the same effect, but in each individual
measurement the effects are drastically different. Most importantly, using
the approach of trajectories we simulate actual experimental realizations,
which is impossible using the master equation.

Since the full quantum trajectory simulation \cite{Wiseman1993,Gambetta2008} of
our system is very difficult computationally, we performed the
simulation only in the regimes of Figs.\ \ref{fig:decay}(a) and
\ref{fig:decay}(c), i.e., assuming the ``bad cavity limit'', $\kappa
\gg (\Delta, \Gamma)$. In this case the full simulation can be
replaced with the simple quantum Bayesian approach
\cite{Korotkov1999,Korotkov2014}. For further simplification we
assumed that the resonator is driven practically on resonance,
$|\Delta_{\rm r}\pm\chi |\ll \kappa$, and the setup in Fig.\
\ref{fig:cQED} uses a phase-sensitive amplifier, which amplifies and
outputs the optimal quadrature $I(t)$, carrying information about
the qubit state (the use of a phase-preserving amplifier can be
described by introducing a limited quantum efficiency, $\eta\leq
1/2$).

The simulations have been performed in the standard quantum Bayesian
way
\cite{Korotkov1999,Korotkov2001b,Korotkov2014,Murch2013,Weber2014},
restricted to the two-qubit single-excitation subspace, i.e., we
simulate evolution of the density matrix with elements
$\rho_{10,10}$, $\rho_{01,01}$, and $\rho_{10,01}$, using the bare
basis.
 In brief,
at each (small) time step $\d t$, the unitary evolution due to the
two-qubit Hamiltonian $H_{\rm q}+H_{\rm qq}$ [see Eqs.\ (\ref{eq:H0}) and
(\ref{eq:Hqq})] is interleaved with the evolution due to
measurement, calculated in the following way. First, the value of
the output signal $I(t)$ (averaged over the duration $\d t$) is
picked randomly from the probability distribution
   \begin{align}\label{eq:idist}
  p(I) =& \rho_{10,10} (t)\,\frac{e^{-(I-I_1)^2/2D}}{(2\pi D)^{1/2}}
  + \rho_{01,01} (t)\,\frac{e^{-(I-I_0)^2/2D}}{(2\pi D)^{1/2}} ,
    \end{align}
where $I_1=1$ and $I_0=-1$ correspond to the bare qubit states $\ket{10}$ and $\ket{01}$, and the variance of the Gaussians is $D = \tau /
\d t$ with the distinguishability time $\tau = (2\eta\Gamma_{\rm m})^{-1}$
defined in Eq.~\eqref{eq:tau}. After picking a random value of $I$,
the density matrix is updated using the relations
\begin{align} \label{eq:bayesian}
  &\frac{\rho_{10,10}(t+\d t)}{\rho_{01,01}(t+\d t)} =
  \frac{\rho_{10,10}(t)}{\rho_{01,01}(t)}\,
  \frac{\exp[-(I-I_1)\,\d t/2\tau]}{\exp[-(I-I_0)\, \d t/2\tau]}, \\
  &\frac{\rho_{10,01}(t+\d t)}
  {\sqrt{\rho_{10,10}(t+\d t)\rho_{01,01}(t+\d t)}} =
  \frac{\rho_{10,01}(t)\, e^{-(\Gamma - \eta\Gamma_{\rm m}) \d t}}
  {\sqrt{\rho_{10,10}(t)\rho_{01,01}(t)}}, \nonumber
\end{align}
where $\rho_{10,10}+\rho_{01,01}=1$ and $\Gamma=\Gamma_{\rm m}+\Gamma_{\rm e}$
may include additional environmental dephasing $\Gamma_{\rm e}$. For
clarity, in what follows we assume $\eta = 1$ and $\Gamma =
\Gamma_{\rm m}$. For a sufficiently small $\d t$, this random sampling and
state update procedure approximates continuous stochastic
trajectories for the two-qubit state $\rho(t)$, as well as for the
normalized readout $I(t) = z(t) + \xi(t)$ that tracks the bare
population difference $z(t) = \rho_{10,10}(t) - \rho_{01,01}(t)$, up
to additive white noise $\xi(t)$ with a constant (two-sided)
spectral density $S = \tau$. While the simulations are performed in
the bare basis, the resulting density matrix $\rho$ can be easily
converted into the eigenbasis. In particular, we are interested in
tracking the eigenbasis populations
$P_{\overline{10}}=\rho_{\overline{10},\overline{10}}$ and
$P_{\overline{01}}=\rho_{\overline{01},\overline{01}}$ besides the
bare basis populations $P_{10}=\rho_{10,10}$ and
$P_{01}=\rho_{01,01}$.

\begin{figure}[t]
  \includegraphics[width=0.75\columnwidth]{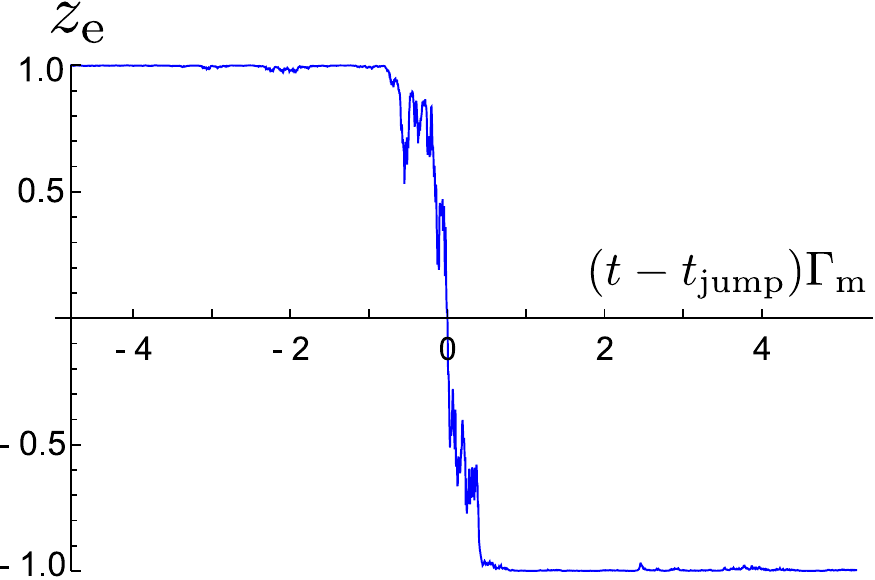}
\caption{(Color online) An example of quantum jump (switching event) between
eigenstates $\ket{\overline{10}}$ and $\ket{\overline{01}}$, obtained in the quantum trajectory simulation for $g/\Delta =1/20$, $\Gamma_{\rm m}/\Delta = 2.5\times 10^{-4}$ and $\kappa \gg \Delta$ (bad cavity regime).
The eigenstate Bloch coordinate $z_{\rm e} = P_{\overline{10}} - P_{\overline{01}}$
noisily hovers near $\pm 1$, except when it rapidly jumps
between the eigenstates on the timescale of $\Gamma_{\rm m}^{-1}$.
Averaging these random jumps produces the decay observed in
Fig.~\ref{fig:decay}(b,c). The physical picture of these jumps is
used later to calculate the measurement error.}
  \label{fig:jump}
\end{figure}

With these simulations, we can analyze transition between the
regimes of Fig.\ \ref{fig:decay}(a) and \ref{fig:decay}(c),
discussed in Sec.\ \ref{sec:eigenbare}. In the textbook regime (a),
$\Gamma_{\rm m}\gg \Delta$, we observe that an initial state (in the
single-excitation subspace) gradually collapses to either the bare
state $\ket{10}$ or $\ket{01}$ within the time scale of (few times)
$\Gamma_{\rm m}^{-1}$, with rare transitions between the bare states at
long time (note that $\eta=1$ and $\kappa\gg \Delta,\Gamma_{\rm m}$).
In contrast, in the regime (c), $\Gamma_{\rm m}\ll
\Delta$, we find from the simulations that the individual
trajectories indeed collapse to the eigenstates
$\ket{\overline{10}}$ and $\ket{\overline{01}}$ at the same time
scale $\Gamma_{\rm m}^{-1}$, as expected from the master equation
simulations. We also observe the expected random \emph{quantum
jumps} between these eigenstates at longer time scales.  An example
of such a quantum jump obtained from the simulations is presented in
Fig.\ \ref{fig:jump}, showing the eigenpopulation difference
$z_{\rm e} = P_{\overline{10}} - P_{\overline{01}}$ switching from
$1$ to $-1$. The typical ``width'' of the jump is comparable to
$\Gamma_{\rm m}^{-1}$, though its central part can be significantly
shorter. In between these random jumps the
states remain close to eigenstates (though sometimes with
``attempts'' of jumps), confirming the assumptions made in the
telegraph noise model of the switching.

Ensemble averaging of the jumps produces the gradual decay of the
population shown in Fig.\ \ref{fig:decay}(c). We have checked
numerically that the averaging of the quantum trajectory results
coincides with the master equation results, thus also
confirming the formula (\ref{eq:Gamma-sw-simple}) for the switching
rate in the regime $\kappa\gg\Delta$. Note that in our trajectory simulations $\Gamma_{\rm sw}^+=\Gamma_{\rm sw}^-$, so the switching can be characterized by a single rate $\Gamma_{\rm sw}$.

We thus numerically confirm our intuitive understanding of the
collapse to the eigenstates and rare switching between them when
$\Gamma_{\rm m}\ll \Delta$. Note that in the simulated regime when also
$\Delta \ll\kappa$, the switching rate is relatively large,
$\Gamma_{\rm sw}\approx 2\Gamma_{\rm m} (g/\Delta)^2$, as follows from
Eq.\ (\ref{eq:Gamma-sw-simple}). This can be understood as because
the fast resonator decay $\kappa$ allows each pump photon to probe
only the \emph{bare states} of the first qubit before escaping to be
collected. In other words, the ``incremental'' measurement
information is still sensitive to the bare basis, even though the
relatively fast interqubit dynamics, $\Delta\gg \Gamma_{\rm m}$, causes
the collapse to occur in the eigenbasis (this is because by the time
``significant'' information is collected, the eigenbasis emerges
as more relevant physically). The tension between the different
bases for the measurement in this ``frustrated'' regime leads to a
relatively large switching rate. In contrast, when $\kappa \ll
\Delta$ (and still $\Gamma_{\rm m}\ll \Delta$), each photon in the
resonator has sufficient time to feel the two-qubit dynamics averaged over the fast
oscillations $\Delta$. Therefore, even the ``incremental''
information in the measurement is sensitive to the eigenbasis, thus
making it very stable and correspondingly reducing the switching
rate $\Gamma_{\rm sw}$. This is a qualitative physical
interpretation of the reduction factor
$\kappa^2/(\kappa^2+4\Delta^2)$ in Eq.\ (\ref{eq:Gamma-sw-simple})
from the point of view  of quantum trajectories.

Note that this interpretation is very different from the physical picture used in our derivation of $\Gamma_{\rm sw}$ in Sec.\ \ref{sec:switching-rate}, in which the reduction factor $\kappa^2/(\kappa^2+4\Delta^2)$ came from non-zero correlation time of the fluctuations $\delta n(t)$. Actually, that picture was based on ``fake'' trajectories for $\delta n(t)$ and was not capable of producing collapse and switching. However, it was capable of describing the ensemble-averaged dynamics, from which we derived $\Gamma_{\rm sw}$ indirectly, by {\it associating} the
ensemble-averaged dynamics with the physically correct picture of quantum jumps. The difference between the two pictures is that quantum trajectories in this section describe actual homodyne measurement, while in Sec.\ \ref{sec:switching-rate} we implicitly assumed a power (photon number) measurement right after the resonator. The two pictures produce the same ensemble-averaged dynamics because of the causality principle, but describe very different evolutions in individual realizations of the measurement.

Since the causality principle is not entirely trivial, let us discuss it in a little more detail. Classical causality requires that an experimenter's action at the present time cannot affect anything in the past. More specifically, the choice of a particular action cannot affect the past. For example, such a choice cannot affect the evolution of an object that has interacted in the past with another object, which is now accessible to the experimenter. However, as we know from the Einstein-Podolsky-Rosen-Bell paradox, this classical causality principle does not work in quantum mechanics, leading in particular to subtle ``delayed choice'' experiments. As a recent example, for a qubit continuously measured in a circuit QED setup, the choice of a measured microwave quadrature (selected with a phase-sensitive amplifier) can dictate the qubit evolution either along meridians or along parallels of the Bloch sphere \cite{Korotkov2014}, even though this choice affects the microwave only after its interaction with the qubit (this prediction has been confirmed experimentally \cite{Murch2013}). Thus, an experimenter's choice in the present may affect the past. However, such passing of information into the past cannot be ``useful'', in the sense that another experimenter in the past cannot extract information about the later choice (otherwise it would be possible to send classical information to yourself in the past; this requirement is often called ``no signaling''). Technically, this limitation is caused by necessarily random results of the measurement: randomness saves causality. In the above example, we can force the qubit retroactively to move along meridians or along parallels, but we cannot control whether the qubit will move right or left (up or down), making it impossible to distinguish the two cases (without using additional information about the microwave measurement result). Because of the no signaling requirement, the choice of the measurement cannot affect the ensemble-averaged evolution in the past (i.e., averaged over the random measurement result in the present), because otherwise it would be possible to extract classical information about the choice. (A similar argument leads to the no-cloning theorem \cite{Wootters1982,Dieks1982}.) Thus, the causality principle in quantum mechanics does not forbid an experimenter to affect evolution in the past; however, the ensemble-averaged evolution in the past (averaged over randomness) cannot be affected by an experimenter's choice (see also \cite{Korotkov2014}).

\section{Qubit measurement error}\label{sec:err}

The switching events (quantum jumps) contribute to the measurement error, which we discuss in this section. Here we consider only the realistic case $\Gamma_{\rm m} \ll \Delta$, when the eigenbasis is preferred for logical encoding over the bare basis. The goal of this section is to find the minimum error, determined by the switching rate $\Gamma_{\rm sw}^-$ (the analysis is very similar to the error limited by the energy relaxation time $T_1$). For simplicity we do not consider transients, assuming that the measurement occurs in the steady state.

We assume that for the readout the information-carrying quadrature $I(t)$ of the output signal is integrated over the measurement duration $t$, producing the averaged output
    \be
    \bar{I}(t)=\frac{1}{t}\int_0^t\!I(t')\,\mathrm{d}t',
    \label{eq:I-bar}\ee
and then this value is compared with the threshold $I_{\rm th}$ to produce a binary readout of ``0'' or ``1.'' (More advanced signal processing of $I(t)$ can moderately improve the measurement fidelity \cite{Gambetta2007,Magesan2015}; we consider the straightforward integration (\ref{eq:I-bar}) for simplicity.)
 If our goal is to distinguish the states $\ket{\overline{10}}$ and $\ket{00}$, then the probabilities of misidentifying these states are
    \be\label{eq:perr1}
  P_{\text{err}}^{(1)} = \int_{-\infty}^{I_{\text{th}}}\!\!\!P(\bar{I}\,|\,\overline{10})\,\d\bar{I},
  \,\,\,\,\,
    P_{\text{err}}^{(0)} = \int_{I_{\text{th}}}^{\infty}\!\!\!P(\bar{I}\,|\,00)\,\d\bar{I},
    \ee
respectively, where $P(\bar{I}\,|\,\overline{10})$ is the probability density of obtaining the result $\bar{I}$ when the initial state is $\ket{\overline{10}}$ and $P(\bar{I}\,|\,00)$ is the analogous probability for the initial state $\ket{00}$. The total measurement error is the average of the two errors,
   \be\label{eq:error-def}
  P_{\text{err}}= \frac{1}{2}\, [ P_{\text{err}}^{(1)} + P_{\text{err}}^{(0)}].
    \ee

\begin{figure}[t]
  \includegraphics[width=0.85\columnwidth]{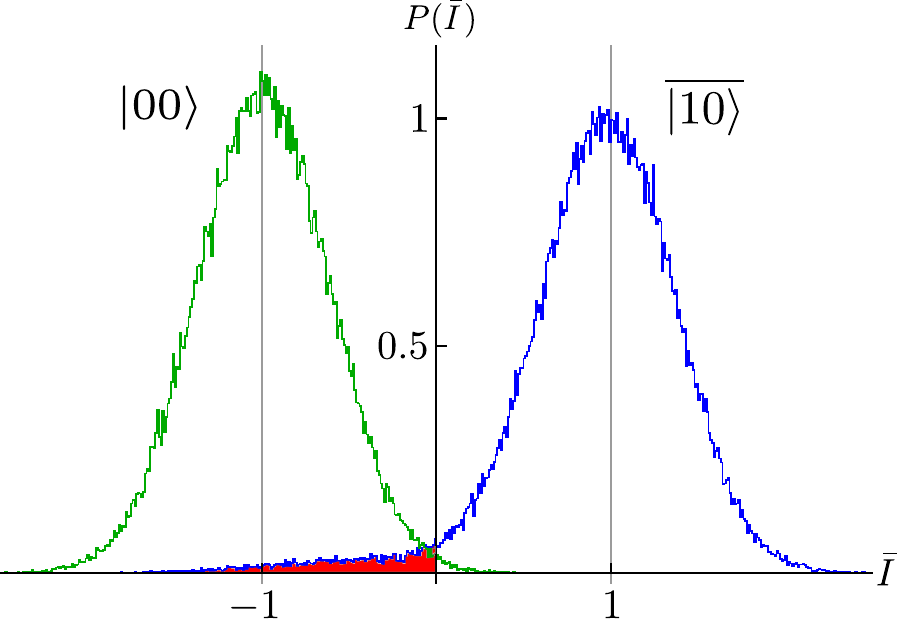}
  \caption{(Color online) Signal histograms $P(\bar{I}\,|\,00)$ and $P(\bar{I}\,|\,\overline{10})$ for the integrated quadrature $\bar{I}$, given the initial logical eigenstates $\ket{00}$ and $\ket{\overline{10}}$. Shown are the binned readouts for $100,000$ trajectories simulated as in Fig.~\ref{fig:jump} for duration $t/\tau = 7$, with $g/\Delta = 1/10$, $\Gamma_{\rm m}/\Delta = 10^{-3}$, and $\eta=1$. Note that $P(\bar{I}\,|\,00)$ is a Gaussian centered at $I_0=-1$, but $P(\bar{I}\,|\,\overline{10})$ is a slightly shifted Gaussian [centered at $1-2(g/\Delta)^2$ instead of $I_1=1$], with a significant extended ``tail'' (red shaded region) caused by quantum jumps. The overlap of the two Gaussians decreases with  integration time $t$, but the histogram overlap due to the tail increases with $t$, thus preventing perfect discrimination. }
  \label{fig:histogram}
\end{figure}

Figure \ref{fig:histogram} shows example histograms for $P(\bar{I}\,|\,00)$ and $P(\bar{I}\,|\,\overline{10})$, obtained by simulating 100,000 quantum trajectories, as discussed in the previous section for $t/\tau =7$, $g/\Delta=1/10$, $\Gamma_{\rm m}/\Delta=10^{-3}$, and $\eta=1$ (in this case $\Gamma_{\rm sw} \tau =9.6\times 10^{-3}$). As in the previous section, we use the normalization in which an ideal single-qubit measurement corresponds to $\bar{I}=I_1=1$ for the state $\ket{1}$ and $\bar{I}=I_0=-1$ for the state $\ket{0}$. As seen from Fig.~\ref{fig:histogram}, the probability distribution $P(\bar{I}\,|\,00)$ is a Gaussian centered at $\bar{I}=-1$, while $P(\bar{I}\,|\,\overline{10})$ has a significant ``tail'' (red shaded region), caused by switching events. Also, the Gaussian part of $P(\bar{I}\,|\,\overline{10})$ is centered at a value slightly smaller than 1 (this shift is practically not visible) because the eigenstate $\ket{\overline{10}}$ has a small contribution from $\ket{01}$. The shape of the histograms is discussed in the appendix.

The errors $P_{\rm err}^{(1)}$, $P_{\rm err}^{(0)}$, and $P_{\rm err}$ depend on the choice of the threshold $I_{\rm th}$. Obviously, the total error $P_{\rm err}$ is minimized when the threshold is set such that $P(I_{\rm th}\,|\,\overline{10})= P(I_{\rm th}\,|\,00)$. However, in most of this section we will assume the symmetric threshold, $I_{\rm th}=0$. This is done for simplicity and also because, as we will see later, the use of the optimal threshold decreases the error insignificantly (with a typical relative improvement of $\alt 3\%$). Also note that we will discuss the error for distinguishing the eigenstates $\ket{\overline{10}}$ and $\ket{00}$ as optimal for logical encoding; the corresponding error for distinguishing the bare states $\ket{10}$ and $\ket{00}$ has an additional contribution,
    \be
    P_{\rm err,bare}^{(1)} \approx  P_{\rm err}^{(1)} + (g/\Delta)^2,
    \ee
because of the initial collapse of the bare state $\ket{10}$ into either the eigenstate $\ket{\overline{10}}$ or $\ket{\overline{01}}$.

\subsection{Error contributions}

In the absence of switching, $\Gamma_{\rm sw}^-=0$, the error steadily decreases with integration time because the variance $\sigma^2=\tau/t$ of the Gaussians in Fig.\ \ref{fig:histogram} decreases with time $t$ [the distinguishability time $\tau$ is given by Eq.\ (\ref{eq:tau})]. In particular, for $I_{\rm th}=0$ this ``separation'' error is
    \begin{eqnarray}\
      && P_{\rm err,sep}^{(0)}= \frac{1}{2}\left[1-{\rm erf}(\sqrt{t/2\tau})\right],      \,\,\,
    \label{eq:separation-error-0} \\
    && P_{\rm err,sep}^{(1)}= \frac{1}{2}\left[1-{\rm erf}(\cos (2\theta)\sqrt{t/2\tau})\right],
    \label{eq:separation-error-1}\end{eqnarray}
where $\cos(2\theta)\approx 1- 2(g/\Delta)^2$ comes from the difference between the eigenbasis and the bare basis. (For the optimal threshold both errors will contain ${\rm erf} [\sqrt{t/2\tau}(1+\cos 2\theta)/2]$.)  For a small $g/\Delta$ this correction is small, and we will neglect it below.

The separation error rapidly becomes very small: $10^{-2}$ for $t=5.4\, \tau$, $10^{-3}$ for $t=9.5\,\tau$, and $10^{-4}$ for $t=13.8\,\tau$. However, the switching process $\ket{\overline{10}}\rightarrow \ket{\overline{01}}$, occurring with the rate $\Gamma_{\rm sw}^-$, adds a contribution to the error $P_{\rm err}^{(1)}$ that increases in time nearly linearly,
    \be
 P_{\rm err}^{(1)} \approx P_{\rm err,sep}^{(1)}
 +\frac{1}{2}\, \Gamma_{\rm sw}^- t,
  \label{eq:err-simple-1}\ee
so that the total error becomes
    \be
    P_{\rm err} \approx \frac{1-{\rm erf}(\sqrt{t/2\tau})}{2} + \frac{1}{4} \, \Gamma_{\rm sw}^- t ,
    \label{eq:err-simple-2}\ee
where we used $\cos (2\theta )\approx 1$, $I_{\rm th}=0$, and $(\Gamma_{\rm sw}^- +\Gamma_{\rm sw}^+) t\ll 1$. More accurate calculations (in particular, taking into account double-switching trajectories, proper convolution of switching and noise, and effects of $\theta$) are presented in the appendix. Note that it is easy to understand the factor $1/2$ in Eq.\ (\ref{eq:err-simple-1}) by saying that the initial state $\ket{\overline{10}}$ will be misidentified only if the switching event occurs before the middle of the integration time, so that the erroneous state is integrated for a longer time than the correct state. (A better interpretation of this factor via symmetry of the convolution is discussed in the appendix.) We also note that Eqs.\ (\ref{eq:err-simple-1}) and (\ref{eq:err-simple-2}) can also describe the error for a single-qubit measurement that accounts for energy relaxation, with $\Gamma_{\rm sw}^-$ replaced by $T_1^{-1}$.

\begin{figure}[t]
    \begin{center}
        \includegraphics[width=0.95\columnwidth]{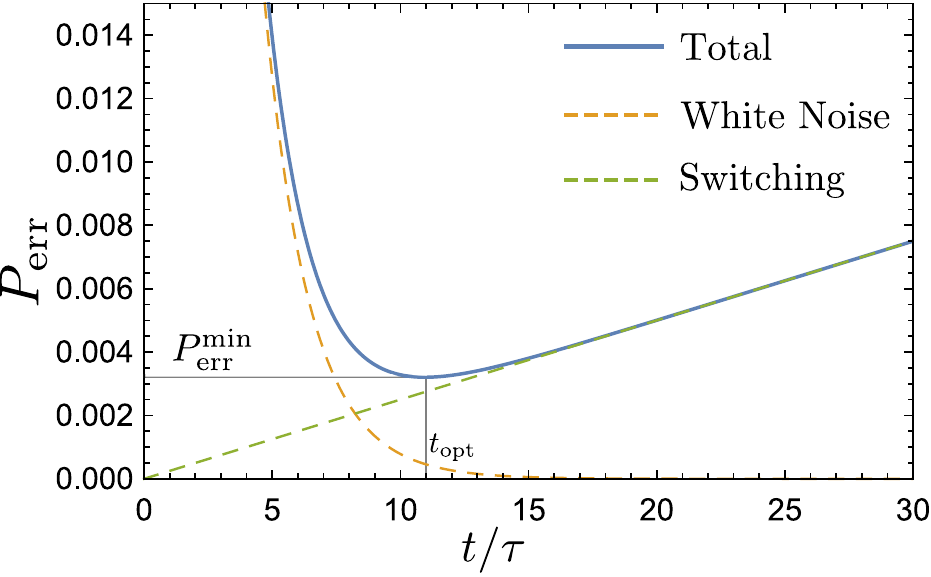}
    \end{center}
\caption{(Color online) Simple analytics for the measurement error $P_{\text{err}}$
as a function of an integration duration $t$, normalized by the
distinguishability time $\tau$. Orange dashed line shows the
monotonically decreasing separation error from integrated white
noise. Green dashed line shows the linearly increasing error
$\Gamma_{\text{sw}}^-t/4$ from switching events, for which we choose
$\Gamma^-_{\text{sw}}\,\tau = 10^{-3}$. The blue solid line shows
the total measurement error, which has a minimum of $P_{\text{err}}^{\rm
min} \approx
(\Gamma_{\text{sw}}^-\tau/2)\,\ln(0.6/\Gamma_{\text{sw}}^-\tau)$ at
the optimum time $t_{\rm opt}/\tau \approx 2 \,\ln
(1/4\Gamma_{\text{sw}}^-\tau)$.
  }
    \label{fig:misid}
\end{figure}

In Fig.~\ref{fig:misid} we illustrate the decomposition of the
eigenbasis measurement error $P_{\rm err}$ of
Eq.~\eqref{eq:err-simple-2} into its two parts for $\Gamma_{\rm
sw}^-\tau  = 10^{-3}$. The orange dashed line shows the error
contribution from the integrated white noise (separation error),
which monotonically decreases with integration time. The green dashed
line shows the error from switching events, which linearly increases
with integration time. The combination of these two opposing effects
in the total measurement error (blue solid line) produces a minimum
error $P_{\rm err}^{\rm min}$ at an optimum time $t_{\rm opt}$,
which we discuss in the following section.

\begin{figure}[t]
    \begin{center}
        \includegraphics[width=0.95\columnwidth]{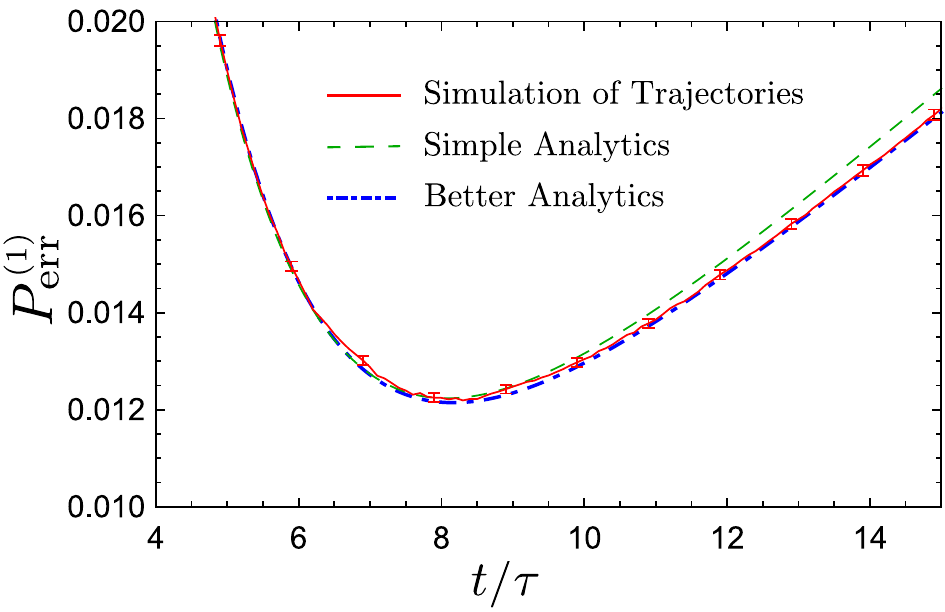}
    \end{center}
\caption{(Color online) Simulated measurement error $P^{(1)}_{\text{err}}$ for
misidentifying the initial state $\ket{\overline{10}}$, using the
discrimination threshold of $I_{\text{th}} = 0$. Red solid line:
measurement error obtained by binning the integrated readouts for
$M=1,500,000$ individual quantum trajectories, with $g/\Delta =
1/20$, $\Delta/\Gamma_{\rm m} = 2000$, and $\eta = 1$, in the bad
cavity regime ($\kappa \gg \Delta$). Error bars show the standard
deviation of $[P_{\rm err}^{(1)}(1-P_{\rm err}^{(1)})/M]^{1/2}$.
Green dashed line: simple analytics that includes only single
quantum jumps, Eq.\ (\ref{eq:err-simple-1}). Blue dot-dashed line:
refined analytics that includes single and double quantum jump
events (see the appendix).
 }
    \label{fig:numerics}
\end{figure}
To verify that this simple approach adequately models the measurement
error, Fig.~\ref{fig:numerics} shows a comparison of the measurement
error $P^{(1)}_{\rm err}$ for the initial state
$\ket{\overline{10}}$ calculated in three ways: using quantum
trajectories, using the simplified description
(\ref{eq:err-simple-1}), and using the more accurate analytics
discussed in the appendix. The quantum trajectory
method has used $M=1,500,000$ individual trajectories initialized in
the eigenstate $\ket{\overline{10}}$ for $g/\Delta = 1/20$,
$\Delta/\Gamma_{\rm m}=2000$, and $\eta = 1$. Each trajectory
consists of $2\times 10^5$ time steps of size $\d t/\tau = 10^{-4}$.
For each trajectory we calculate $\bar I(t)$ via Eq.\
(\ref{eq:I-bar}) and compare it with the threshold $I_{\text{th}} =
0$. The error $P_{\rm err}^{(1)}$ is then the fraction of
trajectories with $\bar{I}<I_{\text{th}}$, which is shown by the
solid red line in Fig.\ \ref{fig:numerics}. Error bars show the
standard deviation $[P_{\rm err}^{(1)}(1-P_{\rm
err}^{(1)})/M]^{1/2}$ for a few representative points. For
comparison, the dashed green line shows the simple analytics
(\ref{eq:err-simple-1})  with $\Gamma_{\rm sw}^-=2\Gamma_{\rm m}
(g/\Omega)^2$ [see Eqs.\ (\ref{eq:Gamma-sw-cl}) and
(\ref{eq:Gamma-sw-simple})], so that $\Gamma_{\rm sw}^-\tau =1/404$. The dot-dashed blue line shows the more
refined analytics described in the appendix that
include double switching events, as well as the proper offset of the
Gaussian by $\cos(2\theta)$. This offset slightly shifts the curve up at times before and near the minimum, while the more accurate account of jumps noticeably shifts the curve down at times after the minimum. [Most of the difference between the green and blue lines at the times after the minimum can be obtained by simply replacing $\Gamma_{\rm sw}^-t$ in Eq.\ (\ref{eq:err-simple-1}) with $1-\exp(-\Gamma_{\rm sw}^-t)$; the remaining difference is mainly due to double switching events.]
As we see, the analytics of the appendix is closer to the
numerical results than the simple analytics, but the difference is
minor. This difference becomes even smaller for smaller ratios
$\Gamma_{\rm sw}^-/\Gamma_{\rm m}$.

\subsection{Error minimized over time}

The simplified model (\ref{eq:err-simple-2}) for the measurement
error has a first contribution that is rapidly decreasing in time,
and a second contribution that is slowly increasing in time (Fig.\
\ref{fig:misid}). Therefore, it has a minimum that is reached at an
optimal time $t_{\rm opt}$. The minimum error $P_{\rm err}^{\rm
min}$ at this optimal time should be determined by the product
$\Gamma_{\rm sw}^-\tau$, since this is the only dimensionless
parameter in the model.  The optimal measurement duration can be
found via the equation $\d P_{\rm err}/\d t=0$, whose solution is a
product-log function, which has the recursive form
    \be\label{eq:tmin-1}
  t_{\text{opt}} = \tau\,
  \ln \frac{2/\pi}{(\Gamma_{\rm sw}^-\tau)^2(t_{\rm opt}/\tau)}.
    \ee
The corresponding minimum error is
    \be\label{eq:pmin-1}
 P_{\text{err}}^{\rm min} \approx  \frac{\Gamma^-_{\text{sw}}\tau }{2}
  + \frac{\Gamma^-_{\text{sw}}t_{\rm opt}}{4} \approx
  \frac{\Gamma^-_{\text{sw}}\tau}{2}\,\ln\frac{C}{\Gamma^-_{\text{sw}}\tau}, \,\,\, C \simeq 0.6,
    \ee
where $C\approx e\sqrt{2/\pi}\sqrt{\tau/t_{\rm opt}}$ actually
depends on $\Gamma^-_{\text{sw}}\tau$, but sufficiently weakly:
$C\in[0.43,\,0.74]$ for $\tau \Gamma^-_{\text{sw}} \in
[10^{-6},\,10^{-2}]$. In deriving the first relation in Eq.\
(\ref{eq:pmin-1}) we used the approximation
    \be
\frac{1-{\rm erf}(x)}{2}\approx \frac{\exp (-x^2)}{2\sqrt{\pi}\, x}, \,\,\, x\gg 1.
    \ee
Note that in the usual case $\tau \Gamma^-_{\text{sw}}\ll 1$, the
main contribution to $P_{\text{err}}^{\rm min}$ comes from the
second term in Eq.\ (\ref{eq:err-simple-2}), with the relative
contribution from the first term being $2\tau/t_{\rm opt}$. The
ratio $t_{\rm opt}/\tau$ can be estimated as $t_{\rm
opt}/\tau\approx 2\,\ln (C/e\Gamma_{\rm sw}^-\tau)\approx 2\, \ln
(1/4\Gamma_{\rm sw}^-\tau)$.

Note that at very long times, $t\agt (\Gamma_{\rm sw}^- +\Gamma_{\rm
sw}^+)^{-1}$, the simplified model in Eq.\ (\ref{eq:err-simple-2})
becomes inapplicable. Moreover, the reverse switching events with
the rate $\Gamma_{\rm sw}^+$ will eventually produce the integrated
output signal $\bar{I}\rightarrow (\Gamma_{\rm sw}^+ - \Gamma_{\rm
sw}^-)/(\Gamma_{\rm sw}^+ + \Gamma_{\rm sw}^-)$ for the initial
state $\ket{\overline{10}}$, which thus can be distinguished with
certainty from the initial state $\ket{00}$ (a similar situation was
discussed for the measurement of phase qubits in Ref.\
\cite{Galiautdinov2012}). However, such long integration times, $t
\gg 1/(\Gamma_{\rm sw}^+ + \Gamma_{\rm sw}^-)$, are impractical even
if we assume the absence of the energy relaxation, so we do not
consider $P_{\rm err}(t)$ for these long times.

\subsection{Optimized threshold}

Let us augment the simplified model (\ref{eq:err-simple-2}) by introducing an arbitrary threshold $I_{\rm th}$; then the error becomes
    \begin{eqnarray}
      && P_{\rm err} \approx \frac{1-{\rm erf}[(1+I_{\rm th})\sqrt{t/2\tau}]}{4}
       \nonumber \\
       && \hspace{0.7cm}+ \frac{1-{\rm erf}[(1-I_{\rm th})\sqrt{t/2\tau}]}{4}
      + \frac{1+I_{\rm th}}{4} \, \Gamma_{\rm sw}^- t . \qquad
    \label{eq:err-simple-3}\end{eqnarray}
Choosing a slightly negative $I_{\rm th}$ decreases the error because of the contribution from the last term (this is also obvious from Fig.\ \ref{fig:histogram} since the histogram for the initial state $\ket{\overline{10}}$ has a long tail).
The optimal threshold $I_{\rm th}^{\rm opt}$ and optimal time $t_{\rm opt}$ can now be found from the system of equations, $\d P_{\rm err}/\d I_{\rm th}=0$ and $\d P_{\rm err}/\d t=0$. These equations are rather lengthy, but in the case $t_{\rm opt}/\tau \gg 1$ lead to a simple relation $\exp (-I_{\rm th} t_{\rm opt}/\tau)=\sqrt{3}$. Therefore, the optimal threshold is only slightly different from zero
    \be
  I^{\rm opt}_{\rm th} \approx -\frac{\ln 3}{2}\, \frac{\tau}{t_{\rm opt}} \approx  -0.55 \, \frac{\tau}{t_{\text{opt}}},
    \label{eq:I-bar-opt}\ee
while the optimal time $t_{\rm opt}$ does not change significantly compared with
Eq.\ (\ref{eq:tmin-1}).

The optimal threshold  $I^{\rm opt}_{\rm th}$ can also be obtained in the following crude way. Using Eq.\ (\ref{eq:err-simple-3}), let us calculate the first and second derivatives of $P_{\rm err}$ over $I_{\rm th}$ at the point $I_{\rm th}=0$. This is simple and gives
$\d P_{\rm err}/\d I_{\rm th}=\Gamma_{\rm sw}^-t/4$, $\d^2P_{\rm err}/\d I_{\rm th}^2=(2\pi)^{-1/2} (t/\tau)^{3/2}\exp (-t/2\tau)$. Then, assuming a parabolic dependence $P_{\rm err}(I_{\rm th})$, we find the optimal threshold as $I_{\rm th}^{\rm opt}= -(\d P_{\rm err}/\d I_{\rm th})/(\d^2P_{\rm err}/\d I_{\rm th}^2)$, which, also using Eq.\ (\ref{eq:tmin-1}), gives $I^{\rm opt}_{\rm th} =  -\tau /2t_{\text{opt}}$. Therefore, this crude derivation does not reproduce the result (\ref{eq:I-bar-opt}) exactly, but is still quite accurate. The error decrease due to optimization of $I_{\rm th}$ can then be found from the same parabolic approximation as $\delta P_{\rm err}^{\rm min}=(\d P_{\rm err}/\d I_{\rm th})I^{\rm opt}_{\rm th}/2$, which gives
    \be
    P_{\rm err}^{\rm min}(I_{\rm th}= I_{\rm th}^{\rm opt}) -
     P_{\rm err}^{\rm min}(I_{\rm th}= 0) \approx -\frac{\Gamma_{\rm sw}^- \tau}{16}.
    \label{eq:factor16}\ee
Since we do not expect a significant change of $P_{\rm err}$ due to a slight shift of $t_{\rm opt}$ in this double-optimization procedure, we can simply replace the term $(1/2)\,\Gamma_{\rm sw}^- \tau$ in Eq.\ (\ref{eq:pmin-1}) with $(7/16)\,\Gamma_{\rm sw}^- \tau$. Therefore, in this crude derivation the error $P_{\rm err}^{\rm min}$ optimized over both time and threshold is still given by Eq.\ (\ref{eq:pmin-1}) with a modified value of $C$,
    \be
    C_{\rm opt} = e^{-1/8} C \approx 0.88\, C \simeq 0.5.
    \ee
The relative decrease of $P_{\rm err}^{\rm min}$ due to the
threshold optimization is approximately $[8\ln (C/\Gamma_{\rm sw}^-
\tau)]^{-1}$, which is about 3\% for $\Gamma_{\rm sw}^-
\tau=10^{-2}$ and smaller for smaller values of $\Gamma_{\rm
sw}^- \tau$.

By solving the optimization problem numerically over a wide parameter range $\Gamma_{\rm sw}^- \tau \in [10^{-6},10^{-2}]$, we have confirmed
that the threshold optimization changes $C$ [defined via Eq.\ (\ref{eq:pmin-1})] by a nearly constant factor $C_{\rm opt}/C\approx 0.88$,
producing the range $C_{\rm opt} \in [0.37,\,0.65]$.
Correspondingly, this produces a nearly insignificant relative
correction of $[1.0\%,3.2\%]$ in the minimum error $P_{\rm err}^{\rm
min}$ over this parameter range.
The denominator 16 in Eq.\ (\ref{eq:factor16}) in numerical results is found to be close to 15. Similarly, we confirmed that Eq.\ (\ref{eq:I-bar-opt}) is satisfied quite well: instead of the factor 0.55, we numerically find 0.55--0.58. This leads to the numerically optimal threshold $I_{\rm th}$ varying only within $[-0.023,\,-0.081]$ over this same parameter range.

Besides using Eq.\ (\ref{eq:err-simple-3}) for the numerical optimization, we also used a modified equation, in which the second term is multiplied by $\exp(-\Gamma_{\rm sw}^- t)$, and in the third term $\Gamma_{\rm sw}^- t$ is replaced with $1-\exp(-\Gamma_{\rm sw}^- t)$. This practically does not change the above mentioned results, except that it slightly lowers $C$: for the same parameter range $\Gamma_{\rm sw}^- \tau \in [10^{-6},10^{-2}]$ it is $C \in [0.43,\,0.64]$ and correspondingly $C_{\rm opt} \in [0.37,\,0.57]$. Note that the factor 4 in the mentioned above approximation $t_{\rm opt}/\tau \approx 2\, \ln (1/4\Gamma_{\rm sw}^-\tau)$ varies within the range $[6.1,\,3.1]$ for zero threshold and the same parameter range, and within $[5.3,\,2.7]$ for the optimal threshold, increasing $t_{\rm opt}$ by about $0.3 \tau$ compared with the zero-threshold case.

We emphasize that the main result of our analysis of the threshold
optimization is that the optimal threshold $I_{\rm th}^{\rm opt}$ is
close to the symmetric point $I_{\rm th}=0$ and that the benefit of
the optimization in decreasing the measurement error is
insignificant. This justifies the use of $I_{\rm th}=0$ in the
analysis. This also shows that it is meaningful not to perform the
threshold optimization in an experiment and instead use the
symmetric point. Note that this conclusion also applies to the case
of qubit energy relaxation, which is analyzed in the same way.

\subsection{Measurement error due to neighboring qubit}

As discussed above, the minimized measurement error can be approximated as
    \be
     P_{\text{err}}^{\rm min} \approx \frac{\Gamma^-_{\text{sw}}\tau}{2}\,\ln\frac{C}{\Gamma^-_{\text{sw}}\tau},
    \ee
where $C\simeq 0.6$ for the symmetric threshold or  $\simeq 0.5$ for the optimal threshold, $\tau=(2\eta\Gamma_{\rm m})^{-1}$ is the distinguishability time, and
    \be
\Gamma^-_{\text{sw}} \approx \frac{1}{T_1} + \frac{2g^2}{\Delta^2+4g^2} \, \frac{8\chi^2 \bar{n}_+}{\kappa}\, \frac{\kappa^2}{\kappa^2+4\Delta^2+16g^2}
    \ee
is the switching rate [see Eq.\ (\ref{eq:Gamma-sw-cl})]. Since in
this paper we are interested in the effect of the neighboring qubit,
let us neglect the energy relaxation rate $1/T_1$. Also, let us use
$|g/\Delta | \ll 1$ and assume $|\Delta_{\rm r}\pm \chi| \alt
\kappa$, so that $8\chi^2 \bar{n}_+/\kappa \approx \Gamma_{\rm m}$.
In this case $\Gamma^-_{\text{sw}}$ is given by Eq.\
(\ref{eq:Gamma-sw-simple}) and the measurement error is
    \be
   P_{\text{err}}^{\rm min} \approx \frac{1}{2\eta}\left(\frac{g}{\Delta}\right)^2\!\! \frac{\kappa^2}{\kappa^2+4\Delta^2} \ln\left[C\eta\,\frac{\kappa^2+4\Delta^2}{\kappa^2(g/\Delta)^2} \right] .
    \ee
Note that this is the error for distinguishing the states $\ket{\overline{10}}$ and $\ket{00}$, while the error for distinguishing the bare-basis states $\ket{10}$ and $\ket{00}$ is larger because of the collapse occuring in the eigenbasis when $|\Delta|\gg \Gamma_{\rm m}$,
    \be
       P_{\text{err,bare}}^{\text{min}} \approx \frac{1}{2}\left(\frac{g}{\Delta}\right)^2 + P_{\text{err}}^{\rm min}.
    \ee

As we see, in the ``bad cavity limit'', $\kappa \gg |\Delta|$, the
eigenbasis error, $P_{\rm err}^{\rm min}\approx
(1/2\eta)(g/\Delta)^2 \ln (C\eta \Delta^2/g^2)$, is quite large, for
example, for $g/\Delta=1/10$ and $\eta =0.2$ we obtain $P_{\rm
err}^{\rm min} \approx 6\%$. The bare-basis error
$P_{\text{err,bare}}^{\text{min}}$ is bigger by $0.5\%$, which is
not significant because $P_{\rm err}^{\rm min}$ is so big.

This may look dangerous for the quantum processors based on
superconducting qubits with ``always-on'' interaction between the
neighbors. Fortunately, typical experimental systems do not operate
in this bad cavity limit; in more realistic parameter regimes the
switching process is strongly suppressed and therefore the
measurement error due to the presence of a neighboring qubit is
relatively small. For example, for $\kappa^{-1}=20\,$ns and $\Delta
/2\pi=0.5\,$GHz, the switching rate is approximately
$10^{-4}\,\Gamma_{\rm m}\,(g/\Delta)^2$, so that for
$g/\Delta=1/10$ and $\eta =0.2$ we obtain a very small measurement
error from the neighboring qubit, $P_{\rm err}^{\rm min}\approx
2\times 10^{-5}$. However, the bare-basis error is still
significant, $P_{\text{err,bare}}^{\text{min}}\approx 0.5\%$, which
means that the bare basis is inappropriate for encoding the logical
information. Since the eigenbasis is also beneficial for logic
operations and idling \cite{Galiautdinov2012}, this makes it an
unambiguously optimal choice for encoding quantum information.

Note that since the switching processes are strongly suppressed in
the regime when $\kappa \ll |\Delta |$, the qubit measurement
remains accurate even when a neighboring qubit is detuned only
moderately, $|\Delta /g|\simeq 3$, as long as the eigenstates are
used for encoding. This fact may simplify the design of quantum
processors in which ``frequency crowding'' may present a problem.

The switching between the eigenstates $\ket{\overline{10}}$ and
$\ket{\overline{01}}$ can be observed experimentally. (For this
purpose it is better to use the jumps
$\ket{\overline{01}}\rightarrow \ket{\overline{10}}$, which can be
easily distinguished from energy relaxation events.)  For example,
for $\kappa^{-1}= 10\,$ns, $g/2\pi=30\,$MHz, $\Delta/2\pi=100\,$MHz,
and $\Gamma_{\rm m}/2\pi=20\,$MHz (corresponding to
$|\chi|/2\pi=2\,$MHz and $\bar{n}=10$), the expected switching time
is about $\Gamma_{\rm sw}^{-1}\simeq 10\,\mu$s.

\section{Conclusion}\label{sec:conc}

We have investigated the measurement error of a superconducting transmon
qubit in a circuit QED setup caused by the coupling $g$ to a detuned
neighboring qubit (or a bus resonator), focusing on the effects of the corresponding
``tail'' population $(g/\Delta)^2$. When the ensemble-dephasing rate
due to measurement is much faster than the qubit-qubit detuning, the
system collapses to the bare energy states as one would expect for a
textbook projective measurement. However, in the more physically
relevant regime with the ensemble-dephasing rate much slower than
the detuning, the system instead collapses to the joint qubit-qubit
\emph{eigenstates}, which are also favorable for quantum operations
and idling. As such, these qubit-qubit eigenstates are the most
appropriate states for high-fidelity logical encoding in realistic
parameter regimes.

We have shown that in regime where joint eigenstates are preferred,
the excitation can randomly jump between these eigenstates while the
qubit is being measured. In between these random jumps, the
two-qubit state is practically pinned to one of the eigenstates. We
have derived the rate of the jumps by using a semiclassical model of
fluctuating ac Stark shift. The obtained analytical result for the
switching rate has been confirmed by comparison with numerical
solution of the master equation, for which the ensemble-averaged
jumps lead to a gradual decay of the initial eigenstate population.
The random jumps produce a contribution to the measurement error
probability that increases almost linearly with integration time in
a way qualitatively similar to the error from
energy-decay processes.

The switching rate for these random jumps depends on the relative
magnitude of the resonator decay and the qubit-qubit detuning. For
quickly decaying resonators, the switching rate is significant and
produces the measurement error exceeding $(g/\Delta)^2$ by several
multiples. However, for more slowly decaying
resonators, as is more typical experimentally, the minimized measurement
error becomes essentially negligible for eigenstate encoding, while
the error for bare basis encoding is still significant and exceeds
$\frac{1}{2}(g/\Delta)^2$.

For the purposes of this study, we have used a static threshold for
digitizing the continuous quadrature readout. We note that more
sophisticated discrimination schemes may be able to take advantage
of the additional information contained in the continuous readout to
partially correct for the switching contribution to the measurement
error.  Generalizing our analysis to multiple
neighboring qubits with simultaneous multi-qubit measurement may also be
interesting for future research. Another possible generalization is
the analysis of decoherence for multiqubit states that involve the
neighboring qubit, which are not supposed to be affected by the
measurement, but are actually influenced by the switching dynamics.
We emphasize that the quantum jumps between the eigenstates
predicted in this paper could be measured experimentally using
existing superconducting qubit technology.

\begin{acknowledgments}
The authors would like to thank Eyob Sete, Eric Mlinar, Julian Kelly, and John Martinis for their useful discussions on the matter. We specially thank John Martinis for pointing out importance of the small-$\kappa$ regime.  The research was funded by the Office of the Director of National Intelligence (ODNI), Intelligence Advanced Research Projects Activity (IARPA), through Army Research Office (ARO) Grant No. W911NF-10-1-0334. All statements of fact, opinion, or conclusions contained herein are those of the authors and should not be construed as representing the official views or policies of IARPA, the ODNI, or the US Government. We also acknowledge support from ARO MURI Grant No. W911NF-11-1-0268 and ARO Grant No. W911NF-15-1-0496.
\end{acknowledgments}

\appendix*

\section{Measurement error derivation}\label{sec:apperr}

In this appendix we provide a more complete derivation of the
measurement error, assuming the regime with $\Gamma_{\rm m} \ll
\Delta$, where the two-qubit eigenstates are the optimal logical
states. Our derivation will consist of two parts. First, neglecting
transients for the resonator and assuming abrupt switching events, we will
calculate the histograms $P(\bar{I}\,|\,00)$ and
$P(\bar{I}\,|\,\overline{10})$ for the integrated measurement output $\bar{I} =
\int_0^t I(t')\,\d t' /t$, corresponding to the initial states
$\ket{00}$ and $\ket{\overline{10}}$, respectively  (we use the word ``histogram'' instead of ``probability distribution'' as a shorter term).
Second, we will impose a discrimination threshold $I_{\text{th}}$ on
these histograms to compute the probability of error according to
our definition in Eq.~\eqref{eq:error-def}.

\subsection{Readout histograms}

Our primary assumption for obtaining the readout histograms is that
we can separate the integrated normalized measurement output $\bar{I}$ into two
approximately uncorrelated terms (signal and noise)
\begin{align}\label{eq:appbayes}
  \bar{I}(t) &= \bar{z}_{\rm tot}(t) + \bar{\xi}(t).
\end{align}
The first term, $\bar{z}_{\rm tot} \equiv \bar{z} + \bar{Z}$, is the
total integrated bare-population difference between the ground and
excited states of the main qubit, which includes a part $\bar{z}$ in
the single-excitation subspace, as well as a part $\bar{Z}$ outside
this subspace,
\begin{align}
  \bar{z}(t) &\equiv \frac{1}{t}\int_0^t[P_{10}(t') - P_{01}(t')]\,\d t', \\
  \bar{Z}(t) &\equiv \frac{1}{t}\int_0^t[P_{11}(t') - P_{00}(t')]\,\d t'.
\end{align}
The second term of Eq.~\eqref{eq:appbayes} is integrated zero-mean white noise, which is randomly sampled from a Gaussian distribution of variance $\tau / t$,
\begin{align}\label{eq:appxi}
 P_\xi(\bar{\xi}) = \sqrt{\frac{t}{2\pi\tau}}\,\exp \left(
 -\frac{\bar{\xi}^2 t}{2\tau}\right) .
\end{align}
We note that the assumption of an uncorrelated dynamics of $\xi(t)$
and $z(t)$ is in general not good in the quantum Bayesian approach.
For example, for single-qubit Rabi oscillations, the correlation
between $\xi(t)$ and $z(t)$ is as strong as autocorrelation for
$z(t)$ \cite{Korotkov2001a}. However, for the dynamics with rare switching events this
assumption should be sufficiently good because the correlation
between $\xi(t)$ and $z(t)$ is most important only in the vicinity
of switching events, which occupy a small fraction of the total
integration time. The approximation (\ref{eq:appbayes}) also
neglects transients of duration $\sim \kappa^{-1}$ at the start of
the measurement and near switching events, implying $t\gg
\kappa^{-1}$.

The approximation of Eq.~\eqref{eq:appbayes} permits us to calculate
each histogram for $\bar{I}$ in a simple way as a convolution
between a histogram for the population difference $\bar{z}_{\rm
tot}$ and the Gaussian white noise distribution for $\bar{\xi}$. If
we additionally assume that in the single-excitation subspace the
state is always pinned to an eigenstate, with abrupt jumps between
the eigenstates (in particular, this implies
$\rho_{\overline{01},\overline{10}}=0$), then the histogram for
$\bar{z}$ is determined by the histogram for the \emph{eigenstate}
population difference  $\bar{z}_{\rm e}$,
\begin{align}
 \bar{z}=\cos(2\theta)\, \bar{z}_{\rm e}, \,\,\,
   \bar{z}_{\rm e} &\equiv \frac{1}{t}\int_0^t[P_{\overline{10}}(t')
   - P_{\overline{01}}(t')]\,\d t' ,
\end{align}
via the conversion factor $\cos(2\theta)$ with rotation angle $2\theta = \arctan(2g/\Delta)$. We will now calculate the histograms corresponding to the specific initial populations $P_{00}(0) = 1$ or $P_{\overline{10}}(0) = 1$, which are the optimal logical states for discrimination.

\subsubsection{Ground-state histogram}

An initial ground state $\ket{00}$ remains $\ket{00}$ for an
arbitrarily long time, and the corresponding output signal
also does not change in time, since we assumed the steady state for
the resonator. Therefore, an initial population $P_{00}(0) = 1$
produces the stationary integrated coordinate $\bar{Z} = -1$ (with
$\bar{z} = 0$) and the stationary integrated total population
difference $\bar{z}_{\rm tot} = -1$, which implies a delta-function
histogram $P_z[\bar{z}_{\rm tot}(t)\,|\,00] = \delta(\bar{z}_{\rm
tot} + 1)$. Convolving this histogram with the Gaussian white noise
in Eq.~\eqref{eq:appxi} produces the histogram for the integrated measurement result,
\begin{align}\label{eq:appground}
  P(\bar{I}\,|\,00) &= \sqrt{\frac{t}{2\pi\tau}}\,\exp\left[ -\frac{(\bar{I} + 1)^2 t}{ 2\tau}\right] ,
\end{align}
which is the expected Gaussian distribution of the same width as the noise that is centered at the ground state normalized signal of $I_0 = -1$.

\subsubsection{Excited-state histogram}

An initially excited eigenstate $\ket{\overline{10}}$ will randomly jump to the eigenstate $\ket{\overline{01}}$ at the rate $\Gamma^-_{\rm sw}$, as discussed in Section~\ref{sec:switching}, and may then randomly jump back to the original eigenstate at the rate $\Gamma^+_{\rm sw}$. We assume that these jumps can be treated as instantaneous compared to the integration time $t$, and that we can treat the eigenstates as stationary between these jumps. We also assume that the jumps obey Poissonian statistics and that the average time between the jumps is long compared to typical integration times, $\Gamma^\pm_{\rm sw}t \ll 1$;    therefore it will be sufficient to consider only zero, one, or two possible jumps per integration duration $t$.  The total histogram for the excited state will then be a weighted contribution of histograms with a definite number of jumps
\begin{align}\label{eq:appexcited}
  P(\bar{I}\,|\,\overline{10}) &= p_0 P^{(0)} (\bar{I}|\overline{10}) + p_1 P^{(1)} (\bar{I}|\overline{10}) + p_2P^{(2)}(\bar{I}|\overline{10}).
\end{align}
We compute each of these histograms and their weights separately.
The derivation is significantly easier for the case when $\Gamma_{\rm sw}^+=\Gamma_{\rm sw}^-\equiv\Gamma_{\rm sw}$, so we will be starting the discussion with this case and then discussing more approximate results for unequal switching rates.

For Poissonian jump statistics with equal switching rates, the probability of having $k$ jumps within the measurement duration $t$ is $p_k=e^{-\Gamma_{\rm sw}t}(\Gamma_{\rm sw}t)^k/k!$. Since we consider only $k\leq 2$, we will use
    \be
    p_0=e^{-\Gamma_{\rm sw}t}, \,\, p_2= \frac{(\Gamma_{\rm sw}t)^2}{2}\, e^{-\Gamma_{\rm sw}t}, \, p_1=1-p_0-p_2,
    \label{p-k-1}\ee
where $p_1$ is chosen to adjust normalization because the largest neglected contribution, $k=3$, also has an odd number of jumps, and such choice slightly improves the accuracy for the calculation of the measurement error (though this is not really important).

In the case of unequal switching rates the exact formulas for $p_k$ are quite lengthy, $p_0= e^{-\Gamma_{\rm sw}^-t}$, $p_1=\Gamma_{\rm sw}^- (e^{-\Gamma_{\rm sw}^- t}-e^{-\Gamma_{\rm sw}^+ t})/(\Gamma_{\rm sw}^+ - \Gamma_{\rm sw}^-)$, $p_2=\Gamma_{\rm sw}^-\Gamma_{\rm sw}^+ [e^{-\Gamma_{\rm sw}^+t}+e^{-\Gamma_{\rm sw}^- t}(\Gamma_{\rm sw}^+t-\Gamma_{\rm sw}^- t-1)]/(\Gamma_{\rm sw}^+-\Gamma_{\rm sw}^-)^2$, so we will use the approximation
    \be
p_0=e^{-\Gamma_{\rm sw}^-t}, \,\, p_2\approx \frac{\Gamma_{\rm sw}^- \Gamma_{\rm sw}^+ t^2}{2}, \, p_1=1-p_0-p_2,
    \label{p-k-2}\ee
Note that the next-order approximation for $p_2$ is $p_2\approx \frac{1}{2}\Gamma_{\rm sw}^-\Gamma_{\rm sw}^+t^2[1-(2\Gamma_{\rm sw}^- +\Gamma_{\rm sw}^+)t/3]$.
Also note that to linear order in $t$ (then fully neglecting $p_2$, as in the main text)
    \be
    p_1 \approx \Gamma_{\rm sw}^- t.
    \ee

The excited-state histogram with zero jumps is similar to the ground-state histogram in Eq.~\eqref{eq:appground}.  The initial eigenpopulation remains stationary in this case, corresponding to a stationary eigenpopulation difference $\bar{z}_{\rm e} = 1$, and thus a bare population difference of $\bar{z} = \cos(2\theta)$ and a histogram of
    \be
P^{(0)}_z(\bar{z}_{\rm tot}\,|\,\overline{10}) = \delta[\bar{z}_{\rm tot} - \cos(2\theta)].
    \ee
Convolving this histogram with the Gaussian noise in Eq.~\eqref{eq:appxi} yields the measurement histogram
\begin{align}
  P^{(0)}(\bar{I}\,|\,\overline{10}) &= \sqrt{\frac{t}{2\pi\tau}}\,\exp \left[-\frac{[\bar{I} - \cos(2\theta)]^2t}{ 2\tau} \right] ,
\end{align}
which is a Gaussian distribution similar to the ground-state histogram, but centered at  $\cos(2\theta) \approx 1 - 2(g/\Delta)^2$ that is slightly shifted from the otherwise expected mean of $I_1=+1$ by the coupling to the neighboring qubit. This slight shift was neglected in the main text as small.

\begin{figure}[t]
    \begin{center}
        \includegraphics[width=0.95\columnwidth]{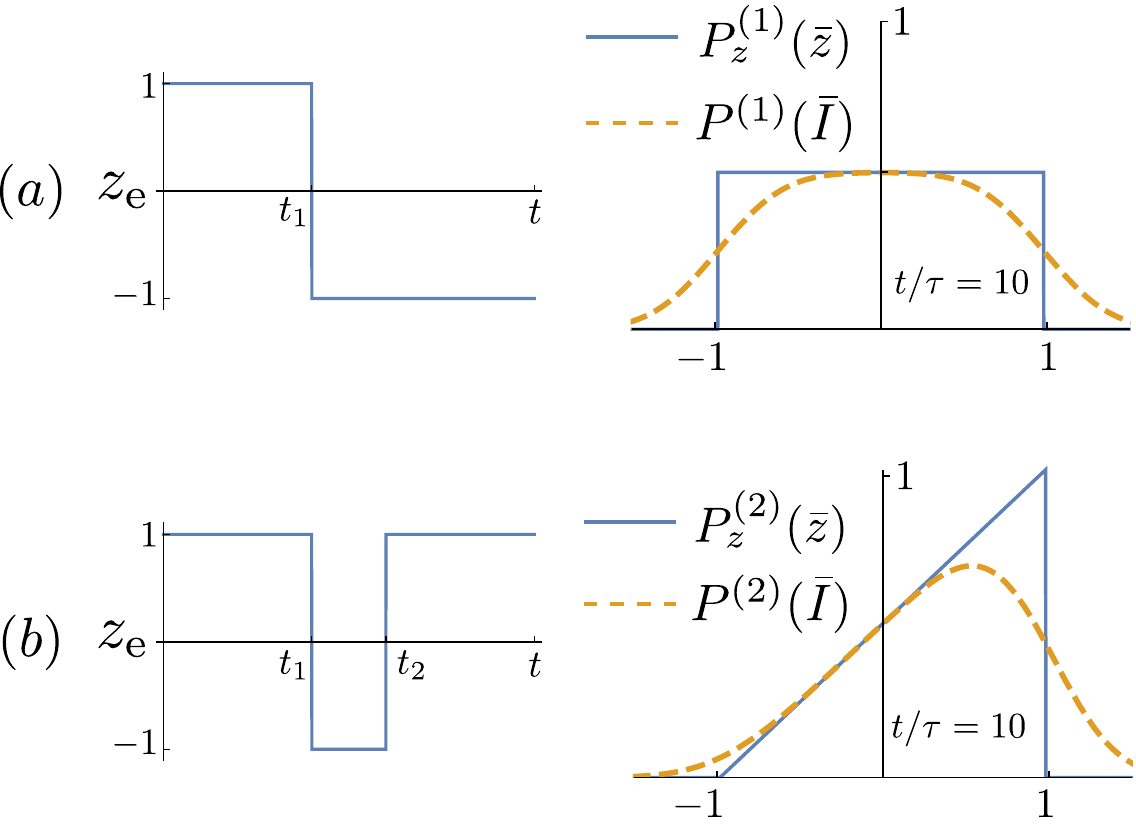}
    \end{center}
    \caption{(Color online) (a) Left panel: schematic evolution of the eigenbasis population difference $z_{\rm e}$ with a jump from $1$ to $-1$ at $t_1$. Right panel: the histogram $P_{z}^{(1)}(\bar{z}_{\rm tot}|\overline{10})$ for time-averaged bare-basis population difference in the one-jump scenario (solid line) and the corresponding histogram $P^{(1)}(\bar{I}|\overline{10})$, which includes Gaussian noise (dashed line).  (b) Similar schematic and histograms for the two-jump scenario. We assume $t/\tau=10$, $g/\Delta =1/10$, and $\Gamma_{\rm sw}^+=\Gamma_{\rm sw}^-$.
    }
    \label{fig:12jumps}
\end{figure}

Now let us calculate the histogram with a {\it single jump}. If the jump occurs at time moment $t_1$, then the signal of $\cos(2\theta)$ is integrated for time $t_1$ and the signal of $-\cos(2\theta)$ is integrated for time $t-t_1$, resulting in the average
    \be \label{eq:zsw1}
  \bar{z}^{(1)} = \frac{2t_1 - t}{t}\,  \cos(2\theta).
    \ee
In the case $\Gamma_{\rm sw}^-=\Gamma_{\rm sw}^+$, the jump time $t_1$ is equally likely at any time in the interval $[0,t]$. Then $\bar{z}_{\rm tot}^{(1)}$  has a uniform histogram $P^{(1)}_z(\bar{z}_{\rm tot}\,|\,\overline{10}) = [2\cos(2\theta)]^{-1}$ in the corresponding interval $[-\cos(2\theta),\,\cos(2\theta)]$, illustrated in Fig.\ \ref{fig:12jumps}(a). Convolving this uniform distribution with the Gaussian white noise in Eq.~\eqref{eq:appxi} yields the measurement histogram
    \be
  P^{(1)}(\bar{I}\,|\,\overline{10}) =\frac{\displaystyle \text{erf}\,\frac{\bar{I} + \cos(2\theta)}{\sqrt{2\tau/t}}
  - \text{erf}\,\frac{\bar{I}-\cos(2\theta)}{\sqrt{2\tau/t}}}{4\cos(2\theta)},
    \label{eq:P(1)(I)}\ee
which is a smoothed box distribution [in Fig.\ \ref{fig:12jumps}(a) the smoothing is shown for $t/\tau=10$]. This addition to the excited-state histogram is the dominant effect of the quantum jumps on the readout. Note that for $t/\tau\agt 4$ the smoothing appreciably affects only the edges of the rectangular distribution  $P^{(1)}_z(\bar{z}_{\rm tot}\,|\,\overline{10})$, so that for $\cos(2\theta)  -|\bar{I}|\agt 2\sqrt{\tau /t}$  we can use $P^{(1)}(\bar{I}\,|\,\overline{10}) \approx P^{(1)}_z(\bar{I}\,|\,\overline{10})$.

In the case of unequal switching rates,  $\Gamma_{\rm sw}^-\neq \Gamma_{\rm sw}^+$, the jump time $t_1$ is no longer equally distributed within $t$; instead, it has the (normalized) probability distribution $e^{-\Gamma_{\rm sw}^-t_1} e^{-\Gamma_{\rm sw}^+ (t-t_1)} (\Gamma_{\rm sw}^+ - \Gamma_{\rm sw}^-)/(e^{-\Gamma_{\rm sw}^-t}-e^{-\Gamma_{\rm sw}^+t})$.  Then the probability distribution for $\bar{z}^{(1)}$ within the interval $[-\cos(2\theta),\,\cos(2\theta)]$ is given by the same formula, with $t_1$ replaced by $[1+\bar{z}^{(1)}/\cos (2\theta )]\,t/2$ and extra normalization factor $t/[2\cos(2\theta)]$. This probability distribution for $\bar{z}^{(1)}$ should then be convolved with the Gaussian noise to obtain $ P^{(1)}(\bar{I}|\overline{10})$. The resulting formula is very long, so for simplicity we can use
    \be
P^{(1)}_z(\bar{z}_{\rm tot}\,|\,\overline{10}) \approx \frac{\displaystyle 1+\frac{(\Gamma_{\rm sw}^+-\Gamma_{\rm sw}^-)t}{2\cos (2\theta)}\, \bar{z}_{\rm tot} }{2\cos(2\theta)}, \,\,\, |\bar{z}_{\rm tot}|\leq \cos(2\theta),
    \label{P-z-1}\ee
and since the convolution of a linear function with the Gaussian noise affects mostly the vicinity of edges, we can use
$P^{(1)}(\bar{I}\,|\,\overline{10}) \approx P^{(1)}_z(\bar{I}\,|\,\overline{10})$ for $\cos(2\theta)  -|\bar{I}|\agt 2\sqrt{\tau/t}$.
A little better approximation is to use Eq.\ (\ref{eq:P(1)(I)}) with added term $\bar{I}(\Gamma_{\rm sw}^+-\Gamma_{\rm sw}^-)t/4\cos^2 (2\theta)$ at $|\bar{I}|\leq \cos(2\theta)$.

Note that if we also want to take into account the energy relaxation with the rate $T_1^{-1}$, then for the energy relaxation event occurring at time $t_1$ we have
$\bar{z}_{\rm tot}^{(1)}=[1+\cos(2\theta)]\,t_1/t -1$. Then using approximation of uniformly distributed $t_1$ (applicable for $t/T_1\ll 1$) we obtain the uniform distribution for $\bar{z}_{\rm tot}^{(1)}$ within the interval $[-1,\,\cos(2\theta)]$. Convolution with the Gaussian noise will then lead to a slightly asymmetric probability distribution $P^{(1)}(\bar{I}|\overline{10})$.

Now let us calculate the histogram with {\it two jumps}. If the first jump occurs at time moment $t_1$ and the return jump occurs at $t_2$, then the system spends duration $\Delta t = t_2-t_1$ in the ``wrong'' state $\ket{\overline{01}}$, so that
    \be\label{eq:zsw2}
  \bar{z}^{(2)}= \frac{t - 2\Delta t}{t}\, \cos(2\theta) .
    \ee
In the case $\Gamma_{\rm sw}^-=\Gamma_{\rm sw}^+$ the probability distribution of time moments $t_1$ and $t_2$ is uniform within the range $0\leq t_1 \leq t_2 \leq t$, and therefore the normalized probability distribution for $\Delta t$  is linearly decreasing, $P_{\Delta t}(\Delta t)=(2/t)[1-(\Delta t)/t]$. This produces the linearly increasing probability distribution for the integrated signal,
    \be
P^{(2)}_z(\bar{z}_{\rm tot}\,|\,\overline{10}) = \frac{\cos(2\theta)+\bar{z}_{\rm tot} }{2\cos^2(2\theta)}, \,\,\, |\bar{z}_{\rm tot}| \leq \cos(2\theta),
    \label{eq:P(2)(z)}\ee
which is illustrated in Fig.\ \ref{fig:12jumps}(b). The convolution with the Gaussian white noise in Eq.~\eqref{eq:appxi} yields the measurement histogram
\begin{align}\label{eq:P(2)(I)}
  P^{(2)}(\bar{I}\,|\,\overline{10}) &= \int_{-\cos(2\theta)}^{\cos(2\theta)}\!\frac{e^{-t(\bar{I}-\bar{z})^2/2\tau}}
  {\sqrt{2\pi\tau/t}}
  \, \frac{\cos(2\theta)+\bar{z}}{2\cos^2(2\theta)} \,\d\bar{z},
\end{align}
which we leave unevaluated here for brevity. As above, we can use approximation $P^{(2)}(\bar{I}|\overline{10})\approx P^{(2)}_z(\bar{I}|\overline{10})$ sufficiently far from the edges, $\cos(2\theta)  -|\bar{I}|\agt 2\sqrt{\tau/t}$ [see Fig.\ \ref{fig:12jumps}(b)].

In the case when $\Gamma_{\rm sw}^-\neq \Gamma_{\rm sw}^+$, the exact formulas are much lengthier because the probability distribution for the jump moments $t_1$ and $t_2$ [which is proportional to $e^{-(\Gamma_{\rm sw}^+ -\Gamma_{\rm sw}^-)(t_2-t_1)}$] is no longer uniform. However, since $|\Gamma_{\rm sw}^+-\Gamma_{\rm sw}^-|\, t \ll 1$, we can still approximate it as a uniform distribution, and then Eqs.\ (\ref{eq:P(2)(z)}) and (\ref{eq:P(2)(I)}) are still approximately valid. Since the two-jump histogram brings a very small contribution to the total histogram (\ref{eq:appexcited}), any crude approximation should be sufficient. Note that if the first jump was due to the energy relaxation event, then the return jump is impossible.

\begin{figure}[t]
    \begin{center}
        \includegraphics[width=0.9\columnwidth]{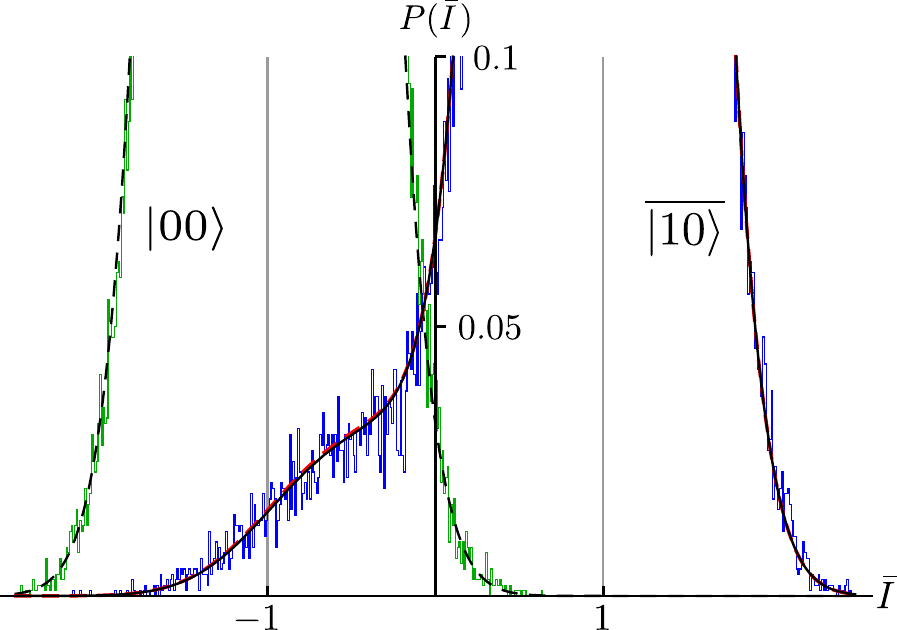}
    \end{center}
    \caption{(Color online) Comparison between analytical and numerical results for the measurement histograms $P(\bar{I}|\,00)$ and $P(\bar{I}|\,\overline{10})$. The green and blue lines show the same numerical results as in Fig.\ \ref{fig:histogram}, but on an enlarged scale. The solid black line shows the analytical result [Eq.\ (\ref{eq:appexcited})] for $P(\bar{I}|\,\overline{10})$, taking into account up to two jumps. The almost coinciding red dashed line shows the result with up to one jump. The black dashed line shows Eq.\ (\ref{eq:appground}) for $P(\bar{I}|\,00)$.
     }
    \label{fig:hist-zoom}
\end{figure}

Thus, the no-jump contribution to the total excited-state histogram (\ref{eq:appexcited}) produces the main Gaussian shape, the single-jump contribution adds an extended nearly uniform tail, and the two-jump contribution produces a very small linearly increasing correction.
   Figure \ref{fig:hist-zoom} shows on an enlarged scale the numerical (quantum trajectory) histograms presented in Fig.\ \ref{fig:histogram} (green and blue lines) and also shows the analytical results derived in this section. The solid black line shows the result for $P(\bar{I}|\overline{10})$ using Eq.\ (\ref{eq:appexcited}) taking into account up to two jumps, with $\Gamma_{\rm sw}^-=\Gamma_{\rm sw}^+=2\Gamma_{\rm m}g^2/(\Delta^2+4g^2)$ for parameters of Fig.\ \ref{fig:histogram} (so that $\Gamma_{\rm sw}^\pm \tau =9.6\times 10^{-3}$ and $t/\tau =7$). The dashed red line shows a similar result using a simplified one-jump approach, in which $p_0= e^{-\Gamma_{\rm sw}^- t}$ and $p_1=1-p_0$.
   We see that the difference between the results for the one-jump and two-jump approaches is very small, but the two-jump approach still agrees slightly better with the numerical results (blue line) for the tail of the distribution. It is interesting to note that the tail looks almost linearly increasing, in contrast to the horizontal shape expected from the uniform distribution of $P^{(1)}_z(\bar{z}_{\rm tot}|\overline{10})$. This is because near its edge, $\bar{I}=-\cos(2\theta)\approx -1$, the Gaussian averaging plays the major role [see Fig.\ \ref{fig:12jumps}(a)]. The black dashed line shows Eq.\ (\ref{eq:appground}) for $P(\bar{I}|00)$; its agreement with the numerical results (green line) is rather trivial because in this case only noise is simulated numerically.

\begin{figure}[t]
    \begin{center}
        \includegraphics[width=0.9\columnwidth]{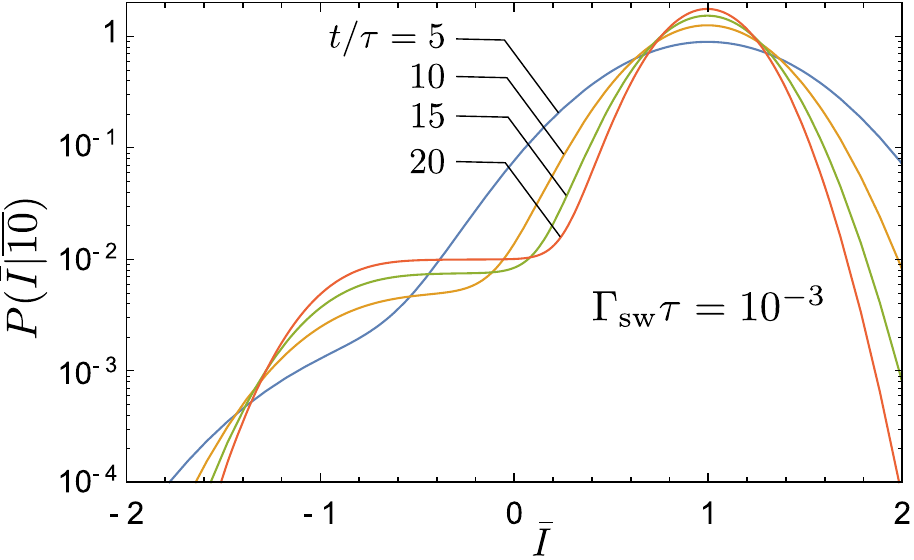}
    \end{center}
    \caption{(Color online) Time evolution of the integrated signal histogram for an initially excited eigenstate $\ket{\overline{10}}$, shown on a semi-log scale, for times $t/\tau = 5,\,10,\,15,\,20$, assuming $\Gamma_{\rm sw}^\pm \tau = 10^{-3}$.  The histogram width due to Gaussian noise decreases with increasing integration time, analogously to the ground state histogram (which is not shown).  Also,  a tail due to quantum jumps between the eigenstates appears at the left side of the histogram; this tail grows in amplitude and flattens with increasing integration time.  }
    \label{fig:hist-evol}
\end{figure}

The evolution of the tail of the excited-state distribution $P(\bar{I}\,|\,\overline{10})$ is illustrated in Fig.~\ref{fig:hist-evol} (we do not show the ground-state histogram for clarity). We assume $\Gamma_{\rm sw}^\pm \tau = 10^{-3}$ and show the histogram at four equally spaced time points, $t/\tau = 5,\,10,\,15,\,20$. At short integration times the Gaussian noise dominates, and the jump events contribute only a small bump distortion in the tail of the Gaussian.
However, as the integration time increases, this bump grows to a flattened tail that becomes significant compared to the otherwise shrinking Gaussian noise. The overlap between $P(\bar{I}\,|\,\overline{10})$ and $P(\bar{I}\,|\,00)$  produces measurement error, which we discuss next.

\subsection{Measurement error probability}

We assume that the states $\ket{00}$ and $\ket{\overline{10}}$ are discriminated by integrating the normalized quadrature $I(t)$ and comparing the result $\bar{I}$ with a threshold value $I_{\rm th}$. Slightly changing the notations used in the main text, we introduce the error probabilities for the two initial states as
    \begin{eqnarray}
&& P_{\text{err}}(t\,|\, 00) = \int_{I_{\rm th}}^{\infty} P(\bar{I}\,|\,00)\,\d\bar{I}, \qquad\qquad\qquad \\
    \,\,\,
&& P_{\text{err}}(t\, |\, \overline{10}) = \int^{I_{\rm th}}_{-\infty} P(\bar{I}\,|\,\overline{10})\,\d\bar{I},
    \label{P-err-10-bar-def}\end{eqnarray}
where in the notations of the main text $P_{\text{err}}(t\,|\, 00)\equiv P_{\rm err}^{(0)}$ and $P_{\text{err}}(t\, |\, \overline{10})\equiv P_{\rm err}^{(1)}$, so that the overall error is
    \be
     P_{\text{err}}(t) = \frac{1}{2}\left[P_{\text{err}}(t\,|\,00) + P_{\text{err}}(t\,|\,\overline{10})\right] .
    \label{P-err-total}\ee

The error for the ground state $\ket{00}$ has a simple form,
    \be
     P_{\text{err}}(t\,|\,00) = \frac{1}{2} \left[1-\mathrm{erf}\,\frac{1 + I_{\rm th}}{\sqrt{2\tau/t}}\right],
    \label{P-err-00}\ee
which follows from Eq.\ (\ref{eq:appground}). The error for the state $\ket{\overline{10}}$ can be calculated as a weighted sum  of  contributions from zero, one, and two jumps, following Eq.\ (\ref{eq:appexcited}),
    \be
P_{\text{err}}(t\,|\,\overline{10}) = p_0 P^{(0)}_{\rm err}(t|\overline{10}) + p_1 P^{(1)}_{\rm err}(t|\overline{10}) + p_2 P^{(2)}_{\rm err}(t|\overline{10}),
    \label{P-err-10-bar}\ee
with the probabilities $p_k$ of $k$ jumps given by Eqs.\
(\ref{p-k-1}) and (\ref{p-k-2}), and the partial error
probabilities,
    \be
P^{(k)}_{\rm err}(t\,|\,\overline{10})=\int_{-\infty}^{I_{\rm th}} P^{(k)}(\bar{I}\,|\,\overline{10})\, \d \bar{I},
    \ee
obtained from the partial histograms $P^{(k)}(\bar{I}\,|\,\overline{10})$ discussed above.

The zero-jump error is then similar to the ground-state error,
    \be
  P^{(0)}_{\text{err}}(t\,|\,\overline{10}) =  \frac{1}{2} \left[1-\mathrm{erf}\,\frac{\cos(2\theta) - I_{\rm th}}{\sqrt{2\tau/t}}\right] ,
    \label{P-err-10-0}\ee
and similarly to Eq.\ (\ref{P-err-00}), it steadily decreases with increasing $t$. The exact formula for the
single-jump error $P^{(1)}_{\text{err}}(t|\overline{10})$ is very
lengthy, but for practical purposes it can be significantly
simplified. First, let us note that in the case $\Gamma_{\rm sw}^-=\Gamma_{\rm sw}^+$,  the distribution
(\ref{eq:P(1)(I)}) for $P^{(1)}(\bar{I}|\overline{10})$ is symmetric
(since it is a convolution of a symmetric distribution
$P^{(1)}_z(\bar{z}_{\rm tot}|\overline{10})$ and Gaussian noise).
Therefore, for the symmetric threshold, $I_{\rm th}=0$, we have
$P^{(1)}_{\text{err}}(t|\overline{10})=1/2$.  Second, for $t/\tau \agt 4$, the distribution
$P^{(1)}(\bar{I}|\overline{10})\approx [2\cos(2\theta)]^{-1}$ is
practically flat for $I_{\rm th}$ close to zero,
$\cos(2\theta)-|I_{\rm th}|\agt \sqrt{4\tau /t}$. Therefore, in this
case $\d  P^{(1)}_{\text{err}}(t|\overline{10})/\d I_{\rm th}
\approx [2\cos(2\theta)]^{-1}$, which gives
    \be
  P^{(1)}_{\text{err}}(t\,|\,\overline{10}) \approx
\frac{1}{2} + \frac{I_{\text{th}}}{2\cos(2\theta)} .
    \label{P-err-1-1}\ee
Very near the symmetric threshold, $|I_{\rm th}|\alt
\tau /3t$, a better approximation is possible,
    \be
  P^{(1)}_{\text{err}}(t\,|\,\overline{10}) \approx
\frac{1}{2} + \frac{I_{\text{th}}}{2\cos(2\theta)} \, \mathrm{erf}\,
\frac{\cos(2\theta)}{\sqrt{2\tau/t}} ,
    \label{P-err-1-2}\ee
which corresponds to the exact derivative at $I_{\rm th}=0$,
$P^{(1)}(\bar{I}=0|\overline{10})= [2\cos(2\theta)]^{-1}  {\rm
erf}[\cos(2\theta)/\sqrt{2\tau/t}]$. Note that very good accuracy for $P^{(1)}_{\text{err}}(t\,|\,\overline{10})$ is not really needed
since its weight $p_1$ in Eq.\ (\ref{P-err-10-bar}) is small.
In the case $\Gamma_{\rm sw}^-\neq \Gamma_{\rm sw}^+$ we can use approximation (\ref{P-z-1}) and neglect the Gaussian averaging, assuming  $\cos(2\theta)-|I_{\rm th}|\agt \sqrt{4\tau /t}$. Then Eq.\ (\ref{P-err-1-1}) generalizes as
    \be
  P^{(1)}_{\text{err}}(t|\overline{10}) \approx
\frac{1}{2} + \frac{I_{\text{th}}}{2\cos(2\theta)} - \frac{(\Gamma_{\rm sw}^+-\Gamma_{\rm sw}^-)t}{8}\left(1-\frac{I_{\rm th}^2}{\cos^2(2\theta)}\right) .
    \label{P-err-1-3}\ee

Note that for $|\Gamma_{\rm sw}^+-\Gamma_{\rm sw}^-|t\ll 1$ and $I_{\rm th}=0$ we have $ P^{(1)}_{\text{err}}(t|\overline{10})\approx 1/2$, which stems from the property that a nearly symmetric distribution (\ref{P-z-1}) for $\bar{z}_{\rm tot}^{(1)}$ remains nearly symmetric after convolution with the Gaussian noise. This explain the factor $1/2$ in Eq.\ (\ref{eq:err-simple-1}) of the main text, which follows from Eq.\ (\ref{P-err-10-bar}) with $p_0\approx 1$, $p_1\approx \Gamma_{\rm sw}^-t$, and $p_2\approx 0$. Similar approximations have been used in Eq.\ (\ref{eq:err-simple-3}), in which we also assumed $\cos(2\theta)\approx 1$.

The double-jump contribution to the total error (\ref{P-err-10-bar}) is very small because of small probability $p_2$. Therefore, it is sufficient to use a crude estimate for $P^{(2)}_{\text{err}}(t|\overline{10})$. In particular, using Eq.\ (\ref{eq:P(2)(z)}) and assuming $P^{(2)}(\bar{I}|\overline{10})\approx P^{(2)}_z(\bar{I}|\overline{10})$, we find
   \be
  P^{(2)}_{\text{err}}(t|\overline{10}) \approx
\frac{1}{4} + \frac{I_{\text{th}}}{2\cos(2\theta)} +  \frac{I_{\text{th}}^2}{4\cos^2(2\theta)} .
    \label{P-err-2}\ee

Even though we neglect the three-jump processes, we actually take into account the main contribution from them automatically, by combining their probability $p_3$ with $p_1$ in Eqs.\ (\ref{p-k-1}) and (\ref{p-k-2}). This is because $P^{(3)}_{\text{err}}(t|\overline{10})\approx 1/2$ for $I_{\rm th}=0$ (following from the symmetry of three-jump processes), which is the same as $P^{(1)}_{\text{err}}(t|\overline{10})$. However, the dependence on $I_{\rm th}$ for the one-jump and three-jump terms is different.

The blue dot-dashed line in Fig.\ \ref{fig:numerics} in the main text shows the error $P_{\rm err}(t\, |\, \overline{10})$ calculated using Eq.\ (\ref{P-err-10-bar}) with the probabilities $p_k$ given by Eq.\ (\ref{p-k-1}), the term $P_{\rm err}^{(0)}(t\, |\, \overline{10})$ given by Eq.\ (\ref{P-err-10-0}), the term $P_{\rm err}^{(1)}(t\, |\, \overline{10})$ equal to $1/2$ (because in Fig.\ \ref{fig:numerics} we use symmetric threshold, $I_{\rm th}=0$), and the term $P_{\rm err}^{(2)}(t\, |\, \overline{10})$ obtained by integration of the histogram (\ref{eq:P(2)(I)}). This analytics fits the numerical result (red solid line in Fig.\ \ref{fig:numerics}) significantly better than the simple analytics (dashed green line) discussed in the main text. Actually, for the two-jump processes it is sufficient to use $P_{\rm err}^{(2)}(t\, |\, \overline{10})=1/4$ [see Eq.\ (\ref{P-err-2})] instead of numerical integration; the result is almost indistinguishable. Note that this is practically equivalent to using $p_0=e^{-\Gamma_{\rm sw}^- t}$, $p_1=\Gamma_{\rm sw}^- t\,[1-(3/4) \Gamma_{\rm sw}^- t]$, $p_2=0$.

Finally, we emphasize that we have used the initial eigenstate $\ket{\overline{10}}$ in the definition for measurement error in Eq.~\eqref{P-err-10-bar-def}, since this is the optimal choice of logical encoding for the regime with $\Gamma_{\rm m} \ll \Delta$. If instead we use the bare state $\ket{10}$, then it will additionally collapse to a mixture of the eigenstates $\ket{\overline{10}}$ and $\ket{\overline{01}}$, which will increase the error,
\begin{align}\label{eq:barereaderr}
& P_{\rm err}(t\,|\,10) =
   \cos^2(\theta)\,P_{\rm err}(t\,|\,\overline{10})
    + \sin^2(\theta)[1-P_{\rm err}(t\,|\,\overline{10})] , \nonumber \\
  &\hspace{1.5cm}= \cos(2\theta)\,P_{\rm err}(t\,|\,\overline{10}) + \sin^2\theta .
\end{align}
Thus, for the total error (\ref{P-err-total}), we will have a nearly constant amount of additional error when distinguishing bare qubit states,  $P_{\text{err,bare}} \approx P_{\text{err}} + (g/\Delta)^2/2$ for $g\ll \Delta$.


%

\end{document}